\RequirePackage{fix-cm}

\documentclass[epjc3]{svjour3}  

\usepackage{hyperref}
\usepackage{graphicx}
\usepackage{amsmath,amssymb}
\usepackage{braket}
\usepackage{multirow}
\usepackage{quantikz}

\usepackage[top=2cm, bottom=2cm, right=2cm, left=2cm]{geometry}

\usepackage[markup=default]{changes}
\definechangesauthor[color=red,name=NunoCastro]{NC}
\definechangesauthor[color=blue,name=GuilhermeMilhano]{GM}

\journalname{Eur. Phys. J. C}
\begin{document}

\title{Jet evolution in a quantum computer: quark and gluon dynamics}

\author{N. F.  Castro\thanksref{e1,dfminho,lipminho}
        \and
        J. G. Milhano\thanksref{e2,liplisboa,ist}
        \and 
M.~G.~Jordão~Oliveira\thanksref{e3,lipminho,nbi} 
}

\thankstext{e1}{e-mail: nuno.castro@cern.ch}
\thankstext{e2}{e-mail: gmilhano@lip.pt}
\thankstext{e3}{e-mail: maria.oliveira@nbi.ku.dk}

\institute{Departamento de F\'{i}sica, Escola de Ci\^{e}ncias, Universidade do Minho, 4710-057 Braga, Portugal \label{dfminho}
           \and
           LIP -- Laborat\'orio de Instrumenta\c{c}\~ao e F\'isica Experimental de Part\'iculas, Escola de Ciências, Campus de Gualtar, Universidade do Minho, 4701-057 Braga, Portugal \label{lipminho}
           \and
           LIP --  Laboratório de Instrumentação e Física Experimental de Partículas, Avenida Professor Gama Pinto 2, 1649-003 Lisboa, Portugal
           \label{liplisboa}
           \and
           Instituto Superior T\'{e}cnico, Universidade de Lisboa, Avenida Rovisco Pais 1, 1609-001, Lisboa, Portugal\label{ist}
           \and
           Present affiliation: NNF Quantum Computing Programme, Niels Bohr Institute, University of Copenhagen, Denmark \label{nbi}
}

\date{\today}

\maketitle

\begin{abstract}
The intrinsic quantum nature of jets and the Quark-Gluon Plasma makes the study of jet quenching a promising candidate to benefit from quantum computing power. Standing as a precursor of the full study of this phenomenon, we study the propagation of SU(3) partons in Quark-Gluon Plasma using quantum simulation algorithms. The algorithms are developed in detail, and the propagation of both quarks and gluons is analysed and compared with analytical expectations. The results, obtained with quantum simulators, demonstrate that the algorithm successfully simulates parton propagation, yielding results consistent with analytical baseline calculations.
\end{abstract}

\section{Introduction}
\label{intro}

Jets, which are highly collimated beams of particles \cite{Salam2010}, are one of the most common probes used to study the Quark-Gluon Plasma (QGP) produced in ultra-relativistic heavy-ion collisions. Jets originate from the fragmentation of hard partons (quarks or gluons). Despite their complexity, jets are excellent probes because their behaviour in vacuum is understood, offering a baseline for comparison. Furthermore, jet properties are modified due to the interaction with the QGP, a phenomenon commonly referred to as jet quenching (for reviews see \cite{Wiedemann:2009sh,Majumder:2010qh,MEHTAR2013,Blaizot2015,Qin:2015srf,Apolinario:2022vzg}). Jet quenching has remained a hot topic throughout the last decades, leading to the development of new techniques to simulate the jet quenching phenomenon and extract QGP properties \cite{Lokhtin2007,Armesto2009,Zapp2014,CasalderreySolana2014,putschke2019}.

Simultaneously, advances in quantum computing have opened new avenues for studying complex systems.  Enabled by the advent of intense sources of highly entangled photons \cite{clauser1972,aspect_1982}, through the exploration of inherent properties of quantum mechanics systems, namely superposition, entanglement, and interference, quantum computers are equipped with operations without classical analogs, enabling the treatment of classical hard or even forbidden problems. Despite all the developments and all the groundbreaking claimed advantages, most of the theoretically proved quantum advantages can only be obtained with fault-tolerant devices which are quite different from the currently available Noise Intermediate-Scale Quantum (NISQ) devices \cite{preskill2018}.
NISQ devices face limitations in qubit count, coherence time, and susceptibility to noise, restricting the practical implementation of quantum algorithms. As a result, the development of quantum software and quantum algorithms is still in a more theoretical stage, with only a few pertinent quantum algorithms being successfully implemented in real quantum devices. So, at present, one may approach the development of quantum algorithms via two main lines of thought: the development of quantum algorithms for the current quantum devices attending to all the limitations and demands, and the development of quantum algorithms for the future fault-tolerant quantum devices without major concerns about the current limitations. Here, with the long-term goal of one day solving the jet quenching problem and noting the the complexity and dimension of the problem, the second line of thought is followed.

In theoretical particle physics, quantum algorithms have been proposed to simulate and solve problems where analytical and numerical solutions are difficult or impossible. One of the most promising applications is simulating quantum field theories using quantum computers \cite{Kreshchuk2022,Jordan2011}, among which there are, for example, applications for parton showers \cite{Spannowsky2022,Bepari2021,Gustafson_2022}, scattering in quantum field theories \cite{Jordan2011,Li_2024}, lattice formulation of QCD \cite{ciavarella2024,bergner2024}, and jet quenching, which is especially relevant for this work. While quantum algorithms for jet quenching are still in their infancy, there has been growing interest in applying quantum computing to this problem. As a precursor of the evolution of a jet in quantum chromodynamics medium, an algorithm for the evolution of a hard probe, a quark, in a colour background was proposed in \cite{Barata2021}. Later, a simpler variant of this algorithm was implemented on quantum simulators and real quantum devices \cite{Barata2022}. This variant was also extended to demonstrate that it can also be used to simulate jet evolution matter. In 2023, these algorithms were further extended to account for gluon production \cite{Barata2023}, and this complex algorithm was implemented in quantum simulators. Recently, Ref. \cite{qian2025} proposed a method for quantum simulation of many-body dynamics using the (3+1)-dimensional QCD Hamiltonian on the light front and presented some results for the SU(2) theory. Here, following Refs. \cite{Li2020,LI2023,Barata2021,Barata2022,Barata2023}, we develop and implement new and more complex quantum algorithms for jet quenching, namely for the evolution of both SU(3) quarks and gluons in QGP media.

\section{Parton in-medium propagation}
\label{sec:Parton_theory}

Jet quenching is due to the interaction of partons in a developing shower with the QGP they traverse.
We consider the propagation of quarks and gluons, referring, whenever appropriate, to them jointly as partons or (hard) probes.

Hard partons propagating through QGP are, owing to their highly boosted kinematics, more conveniently described using light-cone coordinates\footnote{We write the four-momentum of a parton as $p = (p^+,\mathbf{p},p^-)$, where $p^+ = \frac{p^0+p^3}{2}$ corresponds to the light-front energy, $\mathbf{p}$ is the transverse momentum and $p^- = p^0-p^3$. Space-time coordinates are written as $x =(x^+,\mathbf{x},x^-)$, where $x^+ = \frac{x^0+x^3}{2}$ corresponds to the light-cone time, $\mathbf{x}$ is the transverse position and $x^- = x^0-x^3$. Parton propagation is assumed to be along the $\hat{z}$ direction.}.

\begin{figure}
    \centering
    \includegraphics[width = 0.4\textwidth]{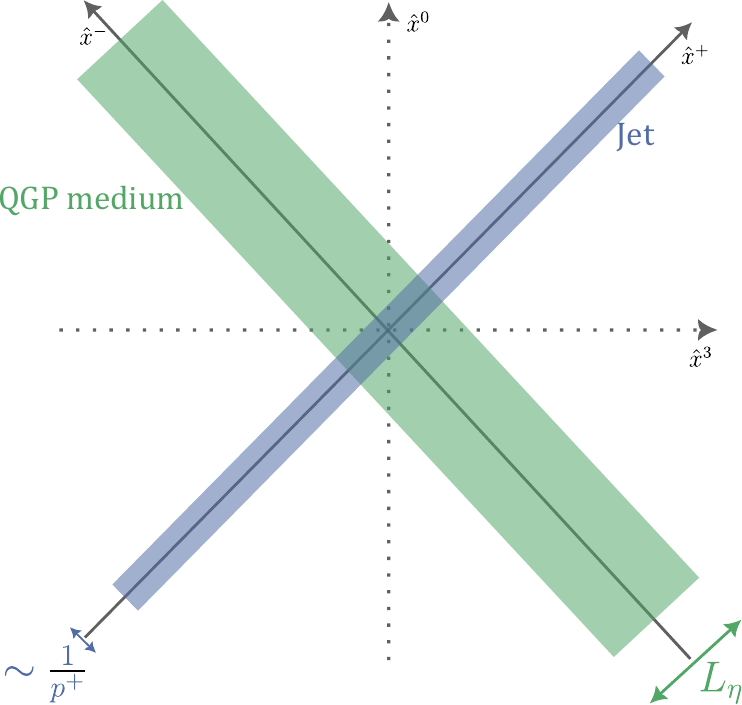}
    \caption[Schematic representation of the jet evolution in a QGP medium.]{Schematic representation of the jet evolution in a QGP medium.}
    \label{fig:medium}
\end{figure}

The QGP is modelled as an external classical stochastic colour field as in \cite{Barata2022}, and further detailed in  \autoref{app:colour_field}.
It is given by $\mathcal{A}^\mu = (\mathcal{A}^+,\mathcal{A}^\perp,\mathcal{A}^-)$, assumed to be boosted along the $\hat{x}^-$ direction, and to have a finite length $L_\eta$ in the $\hat{x}^+$ direction.
A schematic representation of the parton propagation in the QGP under the above conditions is shown in \autoref{fig:medium}.

Since the parton is assumed to be propagating in the $\hat{z}$ direction and that it is highly boosted $(p^z \gg \mathbf{p})$, the dominant component of the parton's light-cone momentum is $p^+$. Furthermore, because the parton is highly localised in $x^-$, the smaller component of the momentum is $p^-$, i.e., $p^+ \gg \mathbf{p} > p^-$. As consequence, in this regime, the spacetime dependence of the field can be simplified to $\mathcal{A}^\mu(x^+,\mathbf{x}, x^-) \approx \mathcal{A}^\mu(x^+,\mathbf{x}, 0) \equiv \mathcal{A}^\mu (x^+,\mathbf{x})$. In the light-cone gauge ($\mathcal{A}^+=0$), the background field can be written as $ \mathcal{A}^\mu = (\mathcal{A}^+,\mathcal{A}^\perp,\mathcal{A}^-) \equiv (\mathcal{A}^\perp,\mathcal{A}^-)$.
Given the smallness of $A^\perp$, the background field $A^\mu$ can be simplified to have $\mathcal{A}^-$ as the only non-zero component, i.e., $\mathcal{A}^\mu\left(x^+,\mathbf{x}\right) \equiv \mathcal{A}^-\left(x^+,\mathbf{x}\right)$. 

Under the sub-eikonal approximation\footnote{The eikonal approximation is usually relaxed to allow small momentum transfers, with the light-cone energy still being conserved, which is known as the sub-eikonal approximation.}, parton propagation in the medium resembles that of a 2D non-relativistic quantum system.  The in-medium scalar propagator of a parton in the transverse plane, from $\left(0,\mathbf{y}\right)$ to $\left(x^+,\mathbf{x}\right)$, is the Green's function $G\left(x^+,\mathbf{x};0,\mathbf{y}\right)$ of the Schrödinger equation \cite{Blaizot2015}
\begin{equation}
    \left(i\partial_{x^+} + \frac{\partial_x^2}{2p^+} + g\mathcal{A}^-(x^+,\mathbf{x})\right)G(x^+,\mathbf{x};0,\mathbf{y}) 
    = i\delta(x^+)\delta(\mathbf{x}-\mathbf{y})\, ,
    \label{eq:schro_propagator}
\end{equation}
where $g$ is a coupling constant. Here, $\mathcal{A}^- (x^+,\mathbf{x} )$ is generically encapsulating both the quark and gluon cases. For the quark case it is defined as 
\begin{equation}
    \mathcal{A}^-(x^+,\mathbf{x}) \equiv \mathcal{A}_a^-(x^+,\mathbf{x})\cdot \mathbf{t}^a\, ,
\end{equation}
where $\mathbf{t} \equiv \frac{\lambda}{2}$ are the SU(3) generators, with $\lambda$ being the Gell-Mann matrices, and $a$ ranging from $1$ to $8$. Similarly, for the gluon case we have

\begin{equation}
    \mathcal{A}^-(x^+,\mathbf{x}) \equiv -i \mathcal{A}_a^-(x^+,\mathbf{x})\cdot {\mathbf{T}^a}^\mathrm{T}\, ,
\end{equation}
where $-i\mathbf{T}$ are the generators in the adjoint representation, and $\mathbf{T}^\mathrm{T}$ are the transpose of $\mathbf{T}$. 

According to \autoref{eq:schro_propagator}, the Hamiltonian that describes the evolution of the parton in the medium is
\begin{equation}
    \hat{H}_{q/g}(x^+) = \frac{\hat{\mathbf{p}}^2}{2p^+} + g\mathcal{A}^-(x^+,\mathbf{x}) = \hat{K} + \hat{V}(x^+)\, .
    \label{eq:hamiltonian}
\end{equation}

The first term $\hat{K} = \frac{\hat{\mathbf{p}}^2}{2p^+}$ corresponds to the kinetic energy of the parton, with the light-cone energy $p^+$ playing the role of mass, and the second term $\hat{V}(x^+) = g\mathcal{A}^-(x^+,\mathbf{x})$ corresponds to the potential energy, which is time-dependent. The Hamiltonian $\hat{H}_{q/g}(x^+)$ rules the evolution of a single parton in the medium, with the subscript $q$ for a quark, the subscript $g$ for a gluon, and the subscript $q/g$ referring to both cases. This Hamiltonian serves as the foundation for developing a quantum algorithm to simulate the parton's in-medium propagation.

\section{Towards parton evolution in a quantum computer}
\label{sec:algorithm}

To study the evolution of a system governed by the Hamiltonian in \autoref{eq:hamiltonian} in a quantum computer, the time evolution operator needs to be defined. In light-cone coordinates, $x^+$ represents the time dimension in which the medium extends over $L_\eta$. Thus, the time evolution operator for the entire medium extension can be written as
\begin{equation}
    \hat{U}\left(L_\eta\right) \equiv \mathcal{T}_+ e ^{-i \int_0^{L_\eta} dx^+ \hat{H}_{q/g}(x^+)}.
    \label{eq:time_evolution}
\end{equation}
%
This operator acts on a two-dimensio\-nal Hilbert space, which is either spanned by the momentum eigenstates $\ket{p}$ or the position eigenstates $\ket{x}$. The time evolution operator evolves the system along the entire longitudinal extension of the medium, transforming the initial state $\ket{\psi_0}$ into a final state $\ket{\psi_{L_\eta}} = \hat{U}\left(L_\eta\right) \ket{\psi_0}$. The operator in \autoref{eq:time_evolution} can be decomposed non-perturbatively into the sequential product of the same operator but for smaller time intervals of size $\Delta_{x^+}$, i.e.,
\begin{equation}
    \hat{U}\left(L_\eta;0\right) = \prod_{j=0}^{N_t-1} \hat{U}\left(x^+_{j+1};x^+_j\right)\, ,
    \label{eq:time_evolution_steps}
\end{equation}
with $x^+_j = j\Delta_{x^+}$, $\Delta_{x^+} = \frac{L_\eta}{N_t}$, and $N_t$ the number of time steps \cite{Barata2021,Barata2022,Li2020,Li2021}. For small enough time steps, in each of them, the time evolution operator can be approximated as the product of the evolution of the kinetic and potential terms. This approximation has corrections of the order of $\Delta_{x^+}^2$,  such that, when $N_t \rightarrow \infty$, $\Delta_{x^+}\rightarrow 0$, it becomes exact with the time evolution operator for the whole medium extension being retrieved. This approximation also facilitates mapping the time evolution operator into a quantum circuit.

To implement the evolution on a quantum computer and retrieve information from the final state $\ket{\psi_{L_\eta}}$, the three major steps of the quantum simulation algorithm \cite{chuang} need to be well-defined as described in the following subsections. The algorithm developed here assumes that the medium probe, the hard parton, is ruled by the SU(3) gauge group,  being either a quark or a gluon. The colour degrees of freedom of both quarks and gluons and their Casimir invariants are summarised in \autoref{table:gauge_infos}. Furthermore, since the kinetic term of the Hamiltonian is diagonal in the momentum basis, and the potential term is diagonal in the position basis, the algorithm is developed in a mixed representation where the momentum and position bases are both used.

\begin{table}
    \centering
    \begin{tabular}{|c|c|c|c|c|}
        \hline
        \multirow{2}{*}{Gauge Group}& \multicolumn{2}{c|}{Colour degrees of freedom} & \multicolumn{2}{c|}{Casimir invariant} \\
        \cline{2-5}
         & Quark & Gluon & $C_F$ & $C_A$ \\
        \hline
        \hline
        SU(3) & $3$ & $8$ & $\frac{4}{3}$ & $3$ \\
        \hline
    \end{tabular}
    \caption[colour degrees of freedom of both quarks and gluons and Casimir invariants for the SU(3) gauge group.]{colour degrees of freedom of both quarks and gluons and Casimir invariants for the SU(3) gauge group. The Casimir invariants are defined as $C_F = \frac{N^2 - 1}{2N}$ and $C_A = N$ for the SU(N) group.}
    \label{table:gauge_infos}
\end{table}

The use of a discrete lattice to simulate the parton propagation in a gauge field can lead to undesired lattice effects in the results. These lattice effects can be avoided by the right choice of some of the simulation parameters. Such as in \cite{Barata2022}, one is interested in avoiding two main effects: (1) ensuring that the lattice spacing captures all the relevant physics -- \textit{spacing effects} -- and (2) ensuring that the assumption that the lattice is finite does not affect the results -- \textit{finite size effects}. For further details on these effects see \autoref{subsec:lattice_effects}.

\subsection{Embedding and initialization}
\label{subsection:embedding_initialization}

The system's data is embedded into quantum states using the basis embedding technique \cite{Barata2022,Li2020}, i.e., the classical bits are directly mapped into quantum bits. For instance, an n-qubit quantum state $\ket{i} = \ket{i_0i_1...i_n}$ represents the classical value $i = i_0i_1...i_n$, where $i_0, i_1, ..., i_n$ are the classical bits. The two-dimensional Hilbert space essentially corresponds to the two dimensions of the transverse lattice, i.e., $\hat{x}$ and $\hat{y}$ directions. Since the algorithm is developed in a mixed position-momentum basis, the change of basis, between the momentum and position bases, is performed using the Quantum Fourier Transform (QFT) \cite{chuang}, which is detailed in \autoref{subsection:encoding_evolution}.

The transverse lattice is a two-dimensional lattice, with $2N_\perp$ sites per direction with each direction extending from $-L_\perp$ to $L_\perp$. The spacing of the lattice is then $\Delta_\perp = \frac{2L_\perp}{2N_\perp} = \frac{L_\perp}{N_\perp}$, in the position basis. Consequently, the lattice spacing in the momentum basis is $\Delta_{\perp p} = \frac{2\pi}{2L_\perp} = \frac{\pi}{L_\perp}$, which corresponds to the reciprocal lattice spacing. The discrete lattice states $\ket{n_x, n_y}$ and $\ket{k_x, k_y}$ are related to the continuum ones $\ket{x, y}$ and $\ket{p_x, p_y}$ by
\begin{equation}
    \ket{x, y} \equiv \Delta_\perp \ket{n_x, n_y}\, , \quad
    n_x, n_y \in \{-N_\perp, -N_\perp + 1, ...,0,..., N_\perp - 1\}\, ,
\end{equation}
and
\begin{equation}
     \ket{p_x, p_y} \equiv \Delta_{\perp p} \ket{k_x, k_y} \quad
     k_x, k_y \in \{-N_\perp, -N_\perp + 1, ...,0,..., N_\perp - 1\}\, .
\end{equation}

A periodic boundary condition is imposed on the lattice, with points separated by an integer multiple of the period $2L_\perp$ being identified:

\begin{equation}
    \hat{O}\ket{n_x, n_y} = \hat{O}\ket{n_x + i2N_\perp, n_y+j2N_\perp}\, ,
\end{equation}
and
\begin{equation}
     \hat{O}\ket{k_x, k_y} = \hat{O}\ket{k_x + i2N_\perp, k_y+j2N_\perp}\, ,
\end{equation}
with $i,j \in \mathbb{Z}$ and $\hat{O}$ a generic operator. This periodic condition identifies the states $\ket{n_i, n_j}$ with $n_i, n_j \in \{-N_\perp, -N_\perp + 1, ...,-1\}$ with the ones with $n_i, n_j \in \{N_\perp, N_\perp + 1, ...,2N_\perp -1 \}$, and the same for the momentum basis, allowing us to work in the interval $[0, 2N_\perp - 1]$ for both the position and momentum bases. Working in this interval not only facilitates the basis change through the QFT/QFT$^\dagger$, but also simplifies the basis embedding due to the non-negativity of the indices. Using basis embedding, the system's state 
can simply be written

\begin{equation}
    \ket{n_x, n_y} \equiv \ket{n_{x_0}n_{x_1}...n_{x_{\log_2(2N_\perp)}},n_{y_0}n_{y_1}...n_{y_{\log_2(2N_\perp)}}}\, ,
\end{equation}
and
\begin{equation}
     \ket{k_x, k_y} = \ket{k_{x_0}k_{x_1}...k_{x_{\log_2(2N_\perp)}},k_{y_0}k_{y_1}...k_{y_{\log_2(2N_\perp)}}}\, ,
\end{equation}
where $n_{x_i}, n_{y_i}, k_{x_i}, k_{y_i} \in \{0,1\}$. 
Consequently, to encode the position/momentum information in the quantum state, we need  $n_q \equiv \log_2(2N_\perp)$ qubits per direction. The qubits representing the position/momentum form two registers, one for each direction.

Besides the transverse lattice, the system's state also encodes information about the colour of the parton. The colour of the parton is also embedded using basis embedding, and, for the SU(3) gauge group, when the parton is a quark, there are three possible colours, and so two qubits are needed to encode the colour information, and when the parton is a gluon, three qubits are needed to encode the colour information.
The qubits representing the colour form also a register, the colour register. The system's state is then a tensor product of the transverse lattice state ($\ket{n_x,n_y}$ or $\ket{k_x,k_y}$) and the colour state ($\ket{c}$), i.e., $\ket{n_x, n_y, c}$ or $\ket{k_x, k_y, c}$. 

Regarding the initialization, the transverse lattice is initialised in the zero momentum state, i.e., $\ket{0,0}$, unless initial momentum effects are of interest. The colour of the parton is initialised in a uniform superposition. 

\subsection{Encoding and evolution}
\label{subsection:encoding_evolution}

After establishing how the system's data is embedded in quantum states, the time evolution operator in \autoref{eq:time_evolution} needs to be discretised and translated into a quantum circuit according to the chosen qubit embedding. Attending to \autoref{eq:time_evolution_steps}, one only needs to define the time evolution operator for a small time interval $\Delta x^+$, i.e., $\hat{U}\left(x^+_{j+1};x^+_j\right)$. Then, the whole time evolution operator is simply the product of these operators. In the context of quantum circuits, this means that one needs to know how to implement the time evolution operator for a small time interval, and then the whole time evolution is simply the sequential application of these operators.

The Hamiltonian defined in \autoref{eq:hamiltonian} has a time dependence only in the potential term, and so the kinetic operator remains the same in the whole evolution. The time dependence of the potential term arises from the background field. Consequently, to include this time dependence, the background field is sliced into $N_\eta$ intervals of size $\Delta_\eta = \frac{L_\eta}{N_\eta}$. In each slice, the background field is assumed to be constant and is generated in a classical computer. Since the medium is assumed to be constant in each slice $\Delta_\eta$, the time steps $\Delta_{x^+}$ should not be larger than $\Delta_\eta$, i.e., $\Delta_{x^+} \leq \Delta_\eta$. For a sufficiently small time step $\Delta_{x^+}$, the time evolution operator for each interval $\Delta_{x^+}$ can be approximated as the product of the evolution of the kinetic and potential terms, and thus written as
\begin{equation}
     \hat{U}\left(x^+_{j+1};x^+_j\right) = \hat{U}\left(x^+_{j}+\Delta_{x^+};x^+_j\right)
     = e^{-i \hat{H}_{q/g}(x^+_j) \Delta_{x^+}} \approx e^{-i \frac{\hat{\mathbf{p}}^2}{2p^+} \Delta_{x^+}} e^{-ig\mathcal{A}^-\left(x^+_j,\mathbf{x}\right)\Delta_{x^+}}.
\end{equation}

When the parton is a quark, this operator is explicitly written as
\begin{equation}
    \hat{U}_q\left(x^+_{j}+\Delta_{x^+};x^+_j\right) \approx e^{-i \frac{\hat{\mathbf{p}}^2}{2p^+} \Delta_{x^+}} e^{-ig\mathcal{A}_a^-\left(x^+_j,\mathbf{x}\right)\mathbf{t}^a\Delta_{x^+}},
\end{equation}
and for a gluon
\begin{equation}
    \hat{U}_g\left(x^+_{j}+\Delta_{x^+};x^+_j\right) \equiv e^{-i \frac{\hat{\mathbf{p}}^2}{2p^+} \Delta_{x^+}} e^{-ig\mathcal{A}_a^-\left(x^+_j,\mathbf{x}\right){\mathbf{T}^a}'\Delta_{x^+}},
\end{equation}
where ${\mathbf{T}^a}' = -i{\mathbf{T}^a}^\mathrm{T}$.

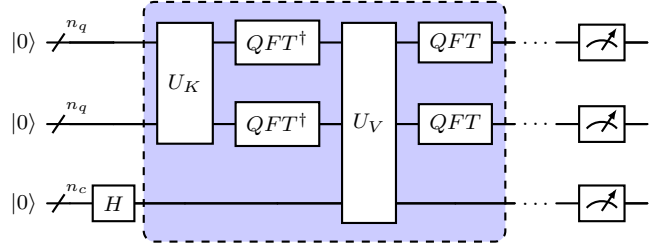
\begin{figure}
    \centering
    \begin{quantikz}[column sep=0.3cm]
        \lstick{$\ket{0}$} & \qwbundle{n_q} & \qw & \gate[2]{U_{K}}\gategroup[3,steps=4,style={dashed,rounded
        corners,fill=blue!20, inner xsep=2pt},background]{} & \gate{QFT^\dagger} & \gate[3]{U_{V}} & \gate{QFT} &  \qw \ \ldots\ &  \meter{} & \qw \\
        \lstick{$\ket{0}$} & \qwbundle{n_q} & \qw & \qw & \gate{QFT^\dagger} & \qw & \gate{QFT} &  \qw  \ \ldots\ & \meter{} & \qw \\ 
        \lstick{$\ket{0}$} & \qwbundle{n_c} & \gate{H} & \qw & \qw & \qw & \qw & \qw  \ \ldots\ & \meter{} & \qw
    \end{quantikz}
    \caption[Quantum circuit for the quantum simulation of a parton in-medium propagation.]{Quantum circuit for the quantum simulation of a parton propagating in a QGP medium, where $n_c$ is the number of colour qubits and $n_q$ is the number of position/momentum qubits per direction. The first step is the initialization, the second step is the application of the time evolution operator, and the third step is the measurement of the final system state. The $U_{K}$ and $U_{V}$ are the kinetic and potential evolution operators, respectively, and the $QFT$ and $QFT^\dagger$ are the gate representation of the QFT and its inverse, respectively. Inside the dashed box, one iteration of the time evolution operator is shown.}
    \label{fig:circuit}
\end{figure}

Since the kinetic term is diagonal in the momentum basis and the potential term is diagonal in the position basis, the implementation of the time evolution operator in each step benefits from a mixed space representation: the kinetic operator is applied in the momentum basis; and, the potential operator is applied in the position basis. Although the operator could be applied in other bases, namely only on the momentum or only on the position basis, the mixed space representation is chosen since it promotes a more straightforward quantum implementation. Within the mixed-state representation, in each time step, the evolution operator corresponds then to the implementation of the kinetic term in the momentum basis, the QFT$^\dagger$ to change to the position basis, followed by the application of the potential term in the position basis, and, finally, the QFT to change back to the momentum basis. 
This framework is summarised as a quantum circuit in \autoref{fig:circuit}, where the dashed box represents one iteration of the time evolution operator, and the initialization and measurement are also included.

The implementation of the QFT and QFT$^\dagger$ follows the framework described in \cite{chuang}. The implementation of the kinetic and potential terms is described separately below.

\subsubsection{Kinetic term}

For each time slice, the time evolution of the kinetic operator is given by
\begin{equation}
    U_K = e^{-i \frac{\hat{\mathbf{p}}^2}{2p^+} \Delta_{x^+}} = e^{-i \frac{\hat{p}_x^2+\hat{p}_y^2}{2p^+} \Delta_{x^+}}.
\end{equation}

The first step is to implement the square of the momentum operator, i.e., $\hat{\mathbf{p}}^2 = \hat{p}_x^2 + \hat{p}_y^2$, in the momentum basis where the momentum operator is diagonal, and then exponentiate it. The effect of applying $\hat{p}^2$ to a momentum state $\ket{p_x,p_y}$ is
\begin{equation}
    \hat{\mathbf{p}}^2 \ket{p_x,p_y} = \left(p_x^2 + p_y^2\right) \ket{p_x,p_y}.
\end{equation}

Knowing that the lattice momentum states $\ket{k_x, k_y}$ are related to the momentum states $\ket{p_x, p_y}$ by $\ket{p_x, p_y} = \Delta_{\perp p}^2 \ket{k_x, k_y}$, the effect of applying $\hat{p}^2$ to a momentum state $\ket{k_x,k_y}$ is
\begin{equation}
    \hat{\mathbf{p}}^2 \ket{k_x,k_y} = \Delta_{\perp p}^2\left(k_x^2 + k_y^2\right) \ket{k_x,k_y}.
\end{equation}

Therefore, the elements of the $\hat{p}^2$ operator in the momentum basis are determined by
\begin{equation}
    \bra{k_x',k_y'} \hat{\mathbf{p}}^2 \ket{k_x,k_y} = \Delta_{\perp p}^2\left(k_x^2 + k_y^2\right) \delta_{k_x',k_x}\delta_{k_y',k_y}.
\end{equation}

Now, one can directly exponentiate the $\hat{p}^2$ operator, i.e., $e^{-i \frac{\hat{\mathbf{p}}^2}{2p^+} \Delta_{x^+}}$ with $p^+$ and $\Delta_{x^+}$ as fixed parameters, and apply the resulting unitary operator to the momentum states. The creation of a unitary gate from a unitary operator/matrix is done using the \texttt{UnitaryGate} class from the Qiskit library \cite{Qiskit2017}. 


\subsubsection{Potential term}
As for the kinetic term, in each time slice, one needs to implement the time evolution operator as a quantum circuit. This operator is given by
\begin{equation}
    \hat{U}_{V_q} = e^{-ig\mathcal{A}_a^-\left(x^+_j,\mathbf{x}\right)\mathbf{t}^a\Delta_{x^+}},
\end{equation}
for a quark, and
\begin{equation}
    \hat{U}_{V_g} = e^{-ig\mathcal{A}_a^-\left(x^+_j,\mathbf{x}\right){\mathbf{T}^a}'\Delta_{x^+}},
\end{equation}
for a gluon. 

The potential term is diagonal in the position basis, and so the implementation of the potential term is carried out on this basis. The first step in the implementation of the potential term is to generate the background field on a classical computer, as described in \autoref{app:colour_field}. Then, the values of the background field need to be contracted with the colour generators for the quark case, or with the structure constants for the gluon case. This contraction can be performed in two different ways, which are both explored and compared. Both methodologies require the representation of the background field as a diagonal matrix in the position basis, denoted as $\hat{\mathcal{A}}^-\left(x^+_j,\mathbf{x}\right)$. When the matrix form of the colour generators has a different shape than $2^{n_c}\times 2^{n_c}$, where $n_c$ is the number of colour qubits, the matrices should be padded with zeros to achieve the correct shape. The same padding rule applies to the ${\mathbf{T}^a}'$ matrices.

The two implementations of colour evolution are as follows:
\begin{itemize}
    \item The first one consists of contracting the background field with the colour generators or structure constants through a tensorial product, i.e., the matrix form of the background field is tensored with the colour generators or the transpose of the matrix form of the structure constants multiplied by $-i$. In this context, being $\hat{\mathcal{A}}_a^-$ the $a^{th}$ component of the background field operator in the matrix form, the contraction with the colour generators is given by $\sum_{a=1}^{N^2-1} \hat{\mathcal{A}}_a^- \otimes \mathbf{t}^a$, and with the matrices ${\mathbf{T}^{a}}'$ is given by $\sum_{a=1}^{N^2-1} \hat{\mathcal{A}}_a^- \otimes {\mathbf{T}^{a}}'$. After computing the sum of tensorial products, the resulting operator can be directly done using the \texttt{UnitaryGate} class from the Qiskit library \cite{Qiskit2017}, or through the decomposition of the operator into a sum of Pauli strings and then exponentiating the Pauli strings. The resulting operator should act in all the qubits, i.e., the position and colour qubits.
    \item The second method starts by inferring that the operator $\hat{U}_V$ can be written as the product of the evolution of each colour component, i.e., $\hat{U}_V = \hat{U}_{V_1} \hat{U}_{V_2} \ldots \hat{U}_{V_{N^2-1}},$ with $\hat{U}_{V_{a_q}} = e^{-ig\hat{\mathcal{A}}_a^- \mathbf{t}^a \Delta_{x^+}}$, for the quark, and $\hat{U}_{V_{a_g}} = e^{-ig\hat{\mathcal{A}}_a^- {\mathbf{T}^{a}}' \Delta_{x^+}}$, for the gluon. Then, the next step consists of verifying that all the matrices $\mathbf{t}^a$ and ${\mathbf{T}^{a}}'$ have eigenvalues $1$ and $-1$. 
    For each colour component, the corresponding colour register should first be diagonalized, i.e., the unitary operator that diagonalizes the colour component is applied to this register. In this diagonal basis, when the colour state is in the eigenstates with eigenvalue $1$, the operator $e^{-ig\hat{\mathcal{A}}_a \Delta_{x^+}}$ should be applied on the position qubits, and when the eigenstate has eigenvalue $-1$, the operator $e^{ig\hat{\mathcal{A}}_a \Delta_{x^+}}$ should be applied instead. After these controlled operations are performed, the colour register is transformed back to its original basis so that the procedure can be repeated for the remaining color components. Finally, in this way, the operators $\hat{U}_{V_a}$ are applied through controlled operations, with the colour qubits being the control qubits, and the position qubits being the target qubits.  As an example, the implementation of the first colour component of a SU(3) quark can be found in \cite{Barata2021}. 
    It should be noted that, even though SU(3) generators do not generally commute, this colour component by colour component application does not introduce significant numerical errors since: (1) the application happens in each small Trotter step of the total evolution, and (2) the Trotter error is proportional to the square of the norm of the background fields, which is small for the regimes not affected by discrete lattice effects. 
    Moreover, this method is easier to implement than the previous one in terms of quantum gates since the background field operators are diagonal in this case, and so easier to describe as a quantum circuit. Consequently, using this approach introduces fewer mapping errors than the previous one, particularly for large problems.
\end{itemize}

\subsection{Measurement and post-processing}
\label{subsection:measurement_post_processing}

The last component of the quantum simulation is the measurement. Although one could implement a sophisticated measurement procedure to extract any desired property from the system's final state -- such as the one proposed in \cite{Barata2021} to directly extract the jet quenching parameter $\hat{q}$ --, here we chose to  retrieve the full probability distribution and then post-process that data as in \cite{Barata2022}. The jet quenching parameter $\hat{q}$, which is a very relevant quantity in the context of jet quenching, is physically defined as the rate of the transverse momentum broadening \cite{Iancu2018}.

By measuring all the qubits, one directly retrieves from the quantum computer the probability distribution of the system's final state, i.e., the probability of finding the system in each of the possible states. The lattice registers are measured on the momentum basis. Recalling the chosen embedding scheme (\autoref{subsection:embedding_initialization}), the measured states are in the form $\ket{k_x, k_y, c} = \ket{k_{x_0}...k_{x_{n_q-1}}, k_{y_0}...k_{y_{n_q-1}}, c_0...c_{n_c-1}}$, with $n_c$ being the number of colour qubits. The measured states then need to be post-processed to extract the probability distribution of the squared transverse momentum. This post-processing is done in four major steps:
\begin{enumerate}
    \item The measured states are converted from the binary representation to the decimal representation;
    \item The lattice is recentered, i.e., the values larger than $N_\perp$ are subtracted by $2N_\perp$, and so the values are in the range $[-N_\perp, N_\perp)$ instead of $[0, 2N_\perp)$;
    \item The lattice values are converted to the physical values, with the lattice values being multiplied by $\Delta_{\perp p}$, i.e., $\ket{k_x, k_y} \rightarrow \ket{k_x\Delta_{\perp p}, k_y\Delta_{\perp p}} = \ket{p_x, p_y}$;
    \item The squared transverse momentum is computed, i.e., $p^2 = p_x^2 + p_y^2$.
\end{enumerate}

We note that when the colour register is initialised in a uniform superposition containing states without physical meaning, i.e. for a SU(3) quark where two qubits are used to encode the three possible colours, the spurious states should be removed from the momentum distribution and, consequently, the distribution should be renormalised.

To better compare the simulation data with the analytical results, it is convenient to consider the transverse momentum transferred by the medium to the parton, commonly referred to as the saturation scale $Q_s^2$, which, in the fundamental representation and neglecting logarithmic corrections, is given by
\begin{equation}
    Q_s^2 = C_F \frac{\left(g^2\mu^2\right)^2L_\eta}{2\pi}\, ,
    \label{eq:saturation}
\end{equation}
where $C_F = \frac{N_c^2-1}{2N_c}$ is the fundamental Casimir of the SU($N_c$) group.

Besides the direct plots and analysis of the probability distribution, determining the jet quenching parameter $\hat{q}$ and comparing it with the analytical results is also of interest. The jet quenching parameter $\hat{q}$ can be extracted from the probability distributions for different background fields through
\begin{equation}
    \hat{q} \equiv \frac{\langle\langle \ \langle \hat{\mathbf{p}}^2\left(L_\eta\right)\rangle \ \rangle\rangle-\langle\langle \ \langle \hat{\mathbf{p}}^2\left(0\right)\rangle \ \rangle\rangle}{L_\eta}\, ,
    \label{eq:jet_quenching_par}
\end{equation}
where $\langle\langle \dots \rangle \rangle$ denotes the average over different background field configurations, and $\langle \hat{\mathbf{p}}^2\left(x\right) \rangle$ denotes the probability weighted average of the squared transverse momentum at the light-front time $x$. When one does not want to study the effect of initial transverse momentum, the initial transverse momentum is set to zero, with the jet quenching parameter being simply given by
\begin{equation}
    \hat{q} \equiv \frac{\langle\langle \ \langle \hat{\mathbf{p}}^2\left(L_\eta\right)\rangle \ \rangle\rangle}{L_\eta}.
    \label{eq:qhat_leta}
\end{equation}

The $\hat{q}$ extracted from the simulation is compared with its analytical counterpart, i.e.,

\begin{equation}
    \hat{q}_{q/g} = \frac{g^4\mu^2 C_{F/A}}{4\pi}
    \left(
    \log\left(\frac{1+\frac{\Delta_\perp^2m_g^2}{\pi^2}}{\frac{1}{N_\perp^2}+\frac{\Delta_\perp^2m_g^2}{\pi^2}}\right)
    \right.
    \left.
    -\frac{\Delta_\perp^2m_g^2}{\pi^2}\left(\frac{1}{\frac{1}{N_\perp^2}+\frac{\Delta_\perp^2m_g^2}{\pi^2}}-\frac{1}{1+\frac{\Delta_\perp^2m_g^2}{\pi^2}}\right)\right)\, ,
    \label{eq:qhat}
\end{equation}

as in \cite{Barata2022,LI2023}. A detailed derivation of \autoref{eq:qhat} can be found in \autoref{app:analytical_jet_quenching}.

\section{Results and analysis}

As seen in \autoref{sec:algorithm}, SU(3) parton propagation can be performed using different implementations of the potential term: via tensorial product (\textit{tensorial} method); and via multiple controlled operations (\textit{controlled}). 
Even though all the results presented here concern physically relevant SU(3) partons, in \autoref{app:SU2} we show some equivalent results for the SU(2) case.

For both a SU(3) quark and a SU(3) gluon, one wants to essentially understand the impact of five parameters: $N_{reps}$, $\Delta_\perp$, $N_{\eta}$, $g^2\mu$ and $p^+$\footnote{The impact of $N_\eta$ and $p^+$ on the jet quenching parameter for analogue classical simulations is done in Ref. \cite{LI2023}}. The first one, $N_{reps}$, is the number of subdivisions of each medium slice $N_{\eta}$, i.e., for each longitudinal medium slice in how many time steps the quark is evolved, which is closely related to the time step $\Delta_{x^+}$. The other four parameters were already well discussed.  The full set of parameters explored in the quantum simulations is presented in \autoref{tab:parameters}. The only parameters that were a priori chosen to avoid undesired lattice effects are $L_{\perp}$ and $m_g$.

\begin{table}
    \centering
    \begin{tabular}{|c|c|}
        \hline
        \textbf{Parameter (Units)} & \textbf{Value(s)} \\
        \hline
        \hline
        $N_{\perp}$ & 4, 8, 16 \\
        \hline
        $L_{\perp}$ (GeV\textsuperscript{-1}) & 4.8 \\
        \hline
        $N_{\eta}$ & 4, 8, 16, 32, 64 \\
        \hline
        $L_\eta$ (GeV\textsuperscript{-1}) & 50 \\
        \hline
        $g^2\mu$ (GeV\textsuperscript{$\frac{3}{2}$}) & 0.004, 0.006, 0.008, 0.01, 0.03, 0.05, 0.06, 0.08, 0.1, 0.5, 1, 1.5, 2 \\
        \hline
        $p^+ $ (GeV) & $+\infty$, 200, 100, 50, 5, 1 \\
        \hline
        $N_{\text{reps}}$ & 1, 2, 4 \\
        \hline
        $m_g$ (GeV) & 0.8 \\
        \hline
        $g$ & 1 \\
        \hline
    \end{tabular}
    \caption{List of the parameters explored in the parton's quantum simulations.}
    \label{tab:parameters}
\end{table}


For each study and for each set of parameters, the $\hat{q}$ is calculated from the retrieved momentum distributions, following \autoref{eq:qhat_leta}, and compared to the analytical expectation obtained from \autoref{eq:qhat}. 
This comparison is always made for different values of $p^+$ and $Q_s^2$, which is computed with  \autoref{eq:saturation} from the $g^2\mu$ value. Furthermore, for each set of parameters, three different executions of the quantum circuit are performed, corresponding to different background field configurations. Each point in the plot is the mean of the three individual executions and the error bars are their standard deviation.

The decomposition of the quantum operators in quantum gates is always performed using the \texttt{UnitaryGate} class from Qiskit's framework \cite{Qiskit2017}. This choice is made since the majority of the studies performed here are executed in simulation environments and the decomposition of the operators into a sum of Pauli matrices brings additional time to the simulation. Furthermore, due to the large size and depth of the quantum circuits, obtained using both methods, the results using a real quantum device are expected to be severely affected by the noise and the decoherence of the quantum device, and so the use of a more time-consuming method is not justified.

The quantum simulations are implemented using Qiskit's framework \cite{Qiskit2017}. The quantum circuits are almost always executed using the \texttt{qasm\_simulator} backend, which is a quantum computer simulator, with the number of shots in each execution being 10000. Due to the actual state of development of quantum devices and their limited access, only a few preliminary quantum simulations were executed in a real quantum device. The simulations were performed on \texttt{ibm\_brisbane} device, an \texttt{Eagle r3} processor, and it was found that the noise completely surpasses the signal, leading to meaningless results. Although this result is not desired, it is expected since error mitigation techniques weren't used, and the circuits are large and complex.

\subsection{Quark}

\begin{figure}
    \centering
    \includegraphics[width=0.8\textwidth]{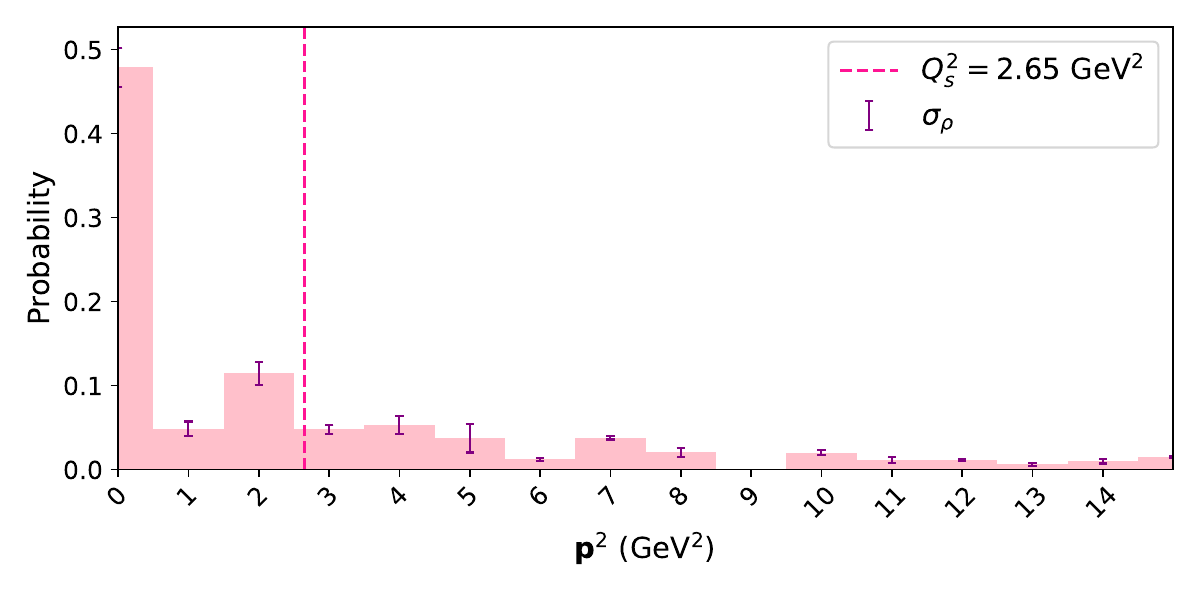}
    \caption{Squared momentum distribution for a SU(3) quark with $N_{\perp} = 8$, $L_{\perp} = 4.8 $~GeV\textsuperscript{-1}, $m_g=0.8$~GeV, $N_{\eta} = 4$, $p^+=+\infty $~GeV, $g^2\mu = 0.5$~GeV\textsuperscript{$\frac{3}{2}$} and $N_{reps} =1 $. For this plot, three individual executions of the quantum circuit are presented, with the height of the bins being the mean of the three individual executions and the error bars being the respective standard deviation. Each execution corresponds to a different background field configuration. The vertical dashed line represents the saturation scale of that simulation.}
    \label{fig:p2quark}
\end{figure}

Although not very informative, the first step to analyse the simulation results is to look at the squared momentum distribution retrieved from the quantum computer. The squared momentum distribution for a SU(3) quark with $N_{\perp} = 8$, $L_{\perp} = 4.8 $~GeV\textsuperscript{-1}, $m_g=0.8$~GeV, $N_{\eta} = 4$, $p^+=+\infty $~GeV, $g^2\mu = 0.5$~GeV\textsuperscript{$\frac{3}{2}$} and $N_{reps} =1 $ is in \autoref{fig:p2quark}. This figure shows the usual tail-shaped momentum distribution, but, apart from that, it isn't easy to compare the results with the analytical expectation and analyse them.

\begin{figure*}
    \centering
    \includegraphics[width=0.9\textwidth]{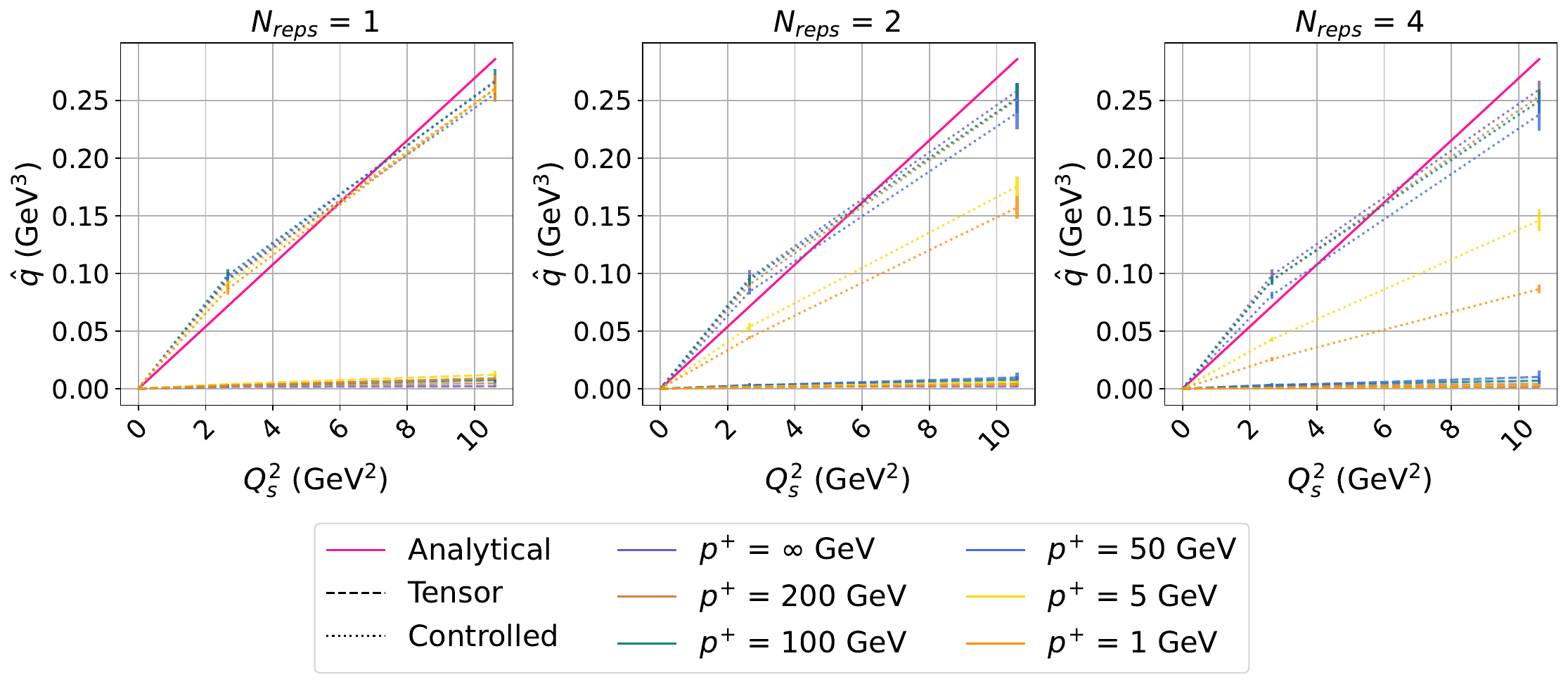}
    \caption{Jet quenching parameter $\hat{q}$ as a function of the saturation scale $Q_s^2$ for a SU(3) quark with $N_{\perp} = 8$, $L_{\perp} = 4.8 $~GeV\textsuperscript{-1}, $m_g=0.8$, $N_{\eta} = 16$ for several values of $p^+$. For each set of parameters, three different executions of the quantum circuit are performed, for different background field configurations, consequently, each point in the plot is the mean of the three individual executions and the error bars are the respective standard deviation. The solid lines represent the analytical expectations.}
    \label{fig:quark_reps}
\end{figure*}

Concerning the impact of the $N_{reps}$ parameter, the same conclusions were always retrieved for the different values of $N_\perp$ analysed. Therefore, we show in \autoref{fig:quark_reps} its impact on the jet quenching parameter for several saturation scales, several values of $p^+$, $N_{reps} \in \{1,2,4\}$ and the two implementation methods, with $N_{\perp} = 8$, $L_{\perp} = 4.8 $~GeV\textsuperscript{-1}, $m_g=0.8$, $N_{\eta} = 16$. 
The analytical expectation \autoref{eq:qhat} is also shown for comparison. 
Even though the increase of the value of $N_{reps}$ is expected to improve the convergence of results, which is the case for larger values of $p^+$, for smaller values deviations coming from the sub-eikonal approximation in \autoref{eq:schro_propagator} are amplified with the increasing of $N_{reps}$. 
To access this error amplification, numerical analysis was done for larger values of $N_{reps}$, which showed that for small $p^+$ values, increasing the just-mentioned parameter leads to tiny $\hat{q}$ values.
Although the smaller $p^+$ values do not correspond to the just-mentioned regime, tests are performed to study the limits and performance across different regimes of the algorithm. 
These deviations for $p^+$ values are consistent with the results reported in Ref. \cite{LI2023}.

Consequently, the value of $N_{reps}$ is fixed to 1 in the other studies. 
Importantly, only the controlled method is in agreement with the analytical expectation. The discrepancy between the methods should be explained by numerical and simulation errors: the tensorial method implies the tensor product of the colour generators matrices by diagonal matrices representing the colour field values, which are small numbers, leading to a large sparse matrix where the non-zero elements are small numbers; then, this matrix is exponentiated, which adds more numerical errors; and, finally, the matrix needs to be converted into quantum gates, which is a process that relies on approximations, introducing more errors. In comparison, the controlled method only requires exponentiating the matrices representing the colour generators component by component and then converting them into quantum gates, which is a much simpler procedure.

\begin{figure*}
    \centering
    \includegraphics[width=0.9\textwidth]{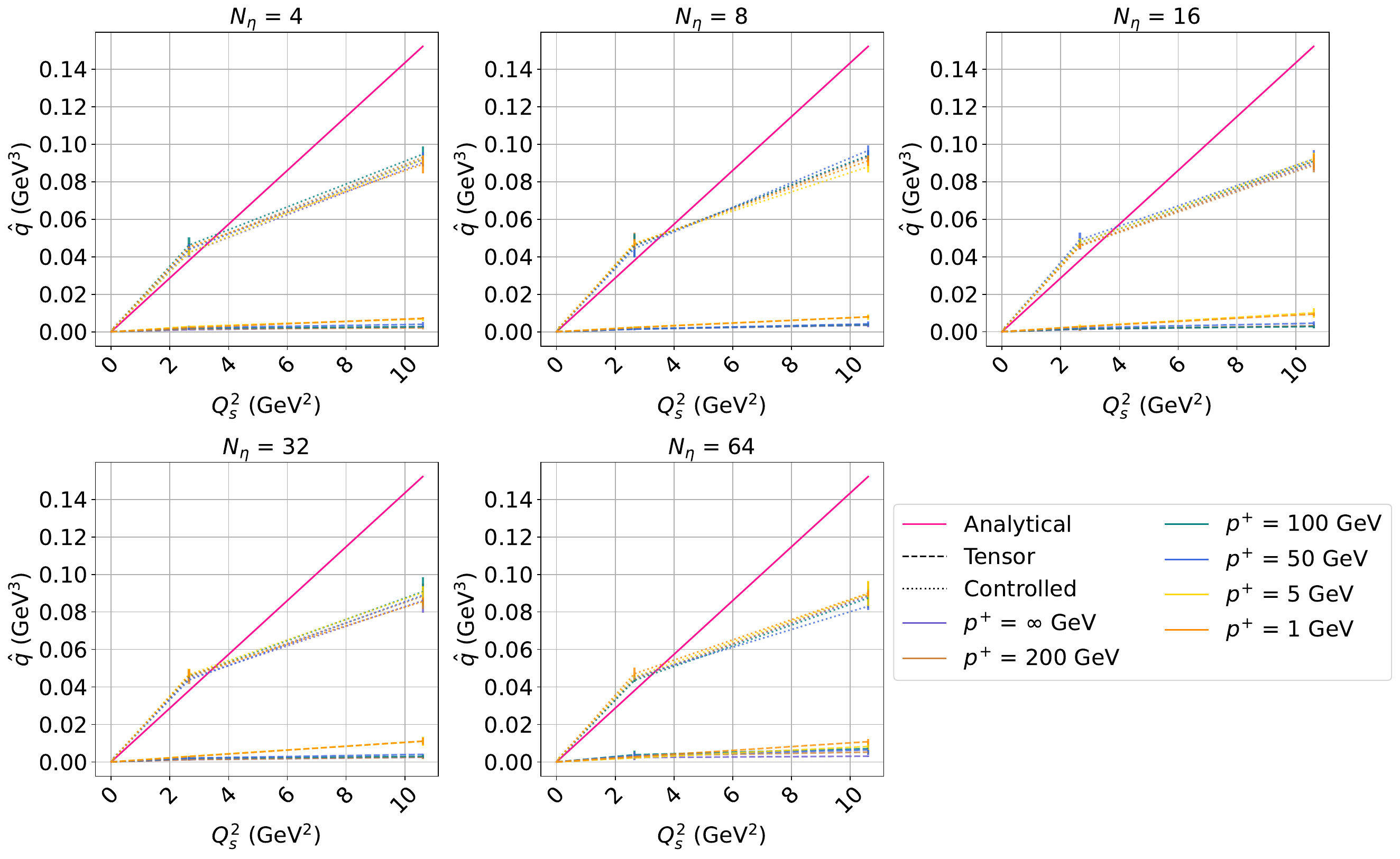}
    \caption{Jet quenching parameter $\hat{q}$ as a function of the saturation scale $Q_s^2$ for a SU(3) quark with $N_{\perp} = 4$, $L_{\perp} = 4.8 $~GeV\textsuperscript{-1}, $N_{reps} = 1$, $m_g=0.8$~GeV for several values of $p^+$ and $N_\eta$. For each set of parameters, three different executions of the quantum circuit are performed, for different background field configurations, consequently, each point in the plot is the mean of the three individual executions and the error bars are the respective standard deviation. The solid lines represent the analytical expectations.}
    \label{fig:quark_neta_nperp4}
\end{figure*}

\begin{figure*}
    \centering
    \includegraphics[width=0.9\textwidth]{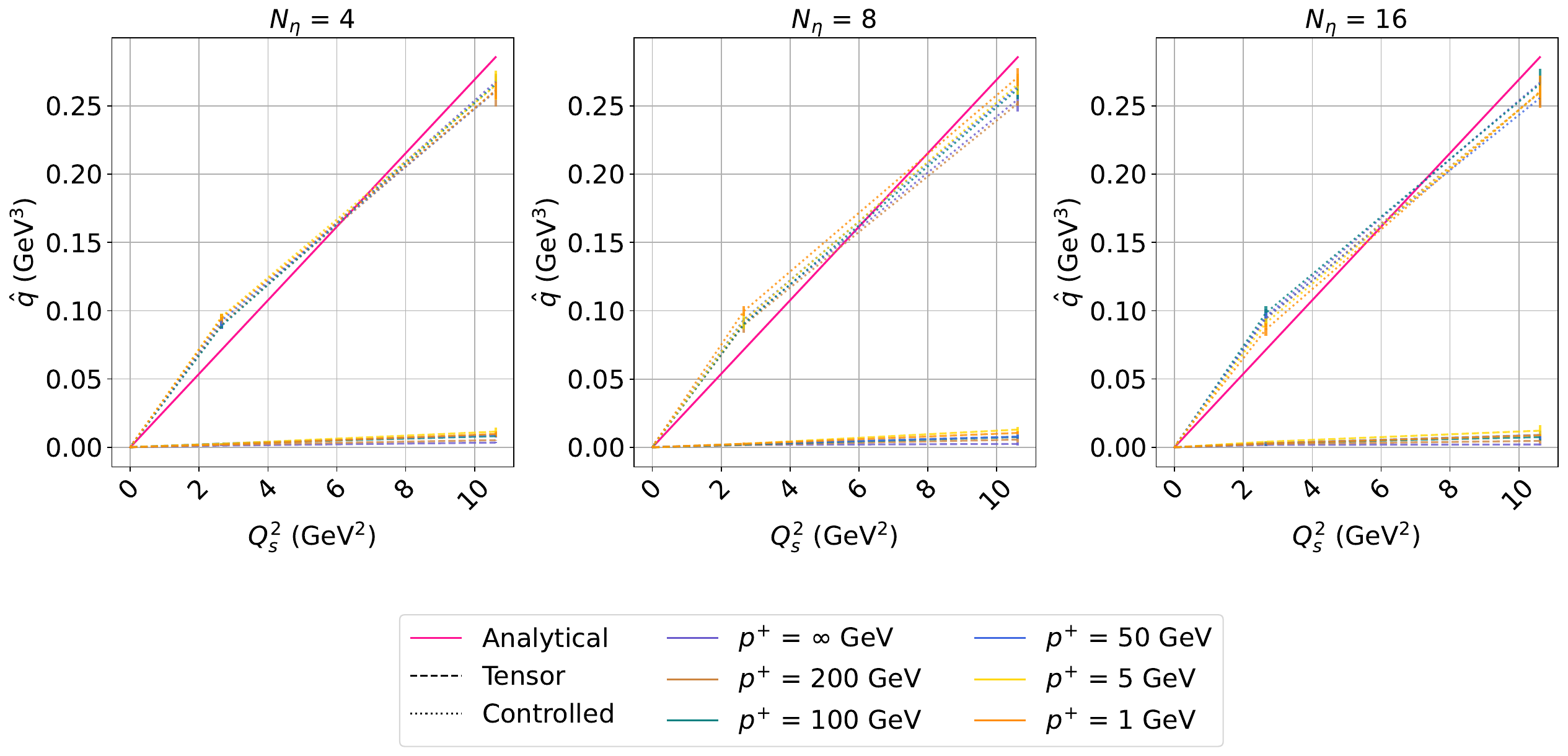}
    \caption{Jet quenching parameter $\hat{q}$ as a function of the saturation scale $Q_s^2$ for a SU(3) quark with $N_{\perp} = 8$, $L_{\perp} = 4.8 $~GeV\textsuperscript{-1}, $N_{reps} = 1$, $m_g=0.8$~GeV for several values of $p^+$ and $N_\eta$. For each set of parameters, three different executions of the quantum circuit are performed, for different background field configurations, consequently, each point in the plot is the mean of the three individual executions and the error bars are the respective standard deviation. The solid lines represent the analytical expectations.}
    \label{fig:quark_neta_nperp8}
\end{figure*}

\begin{figure*}
    \centering
    \includegraphics[width=0.9\textwidth]{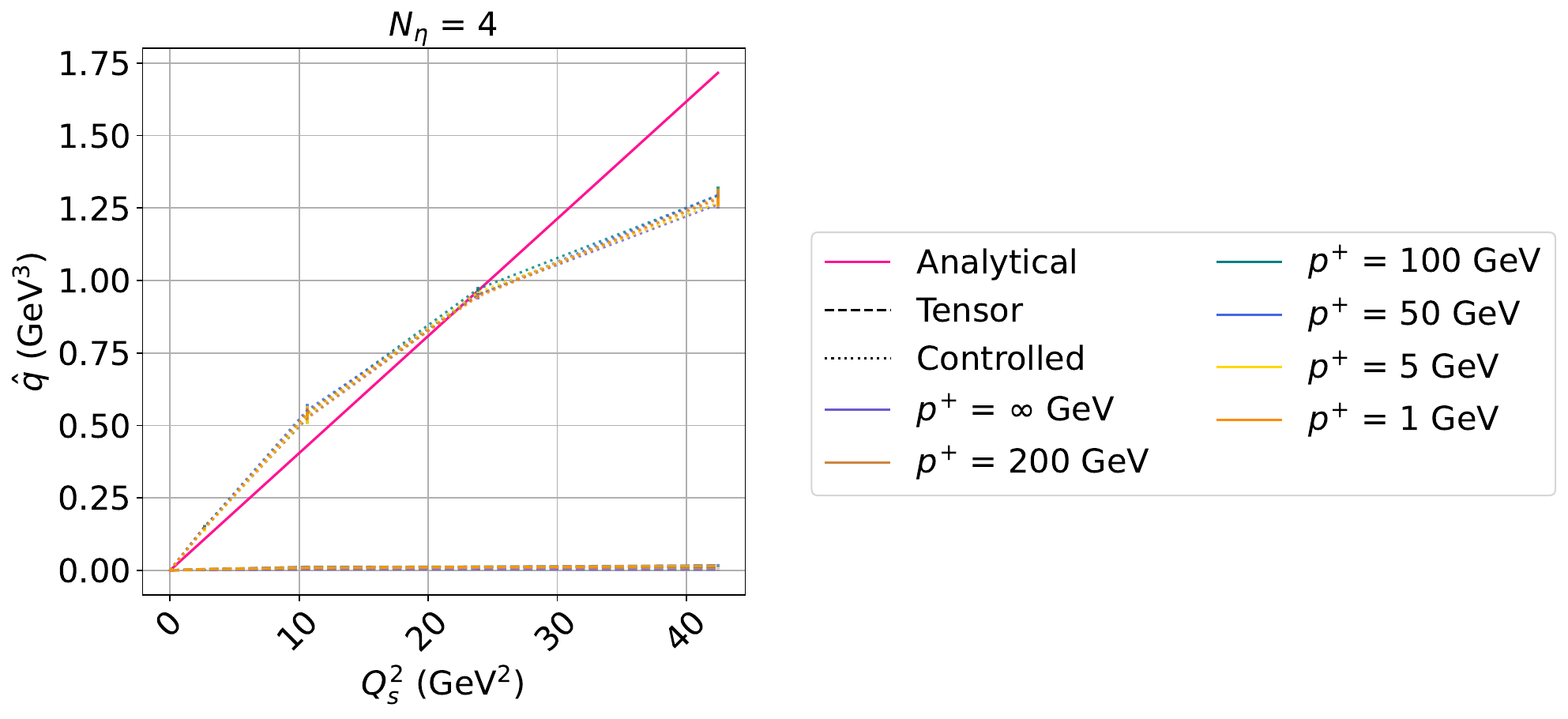}
    \caption{Jet quenching parameter $\hat{q}$ as a function of the saturation scale $Q_s^2$ for a SU(3) quark with $N_{\perp} = 16$, $L_{\perp} = 4.8 $~GeV\textsuperscript{-1}, $N_{reps} = 1$, $m_g=0.8$~GeV for several values of $p^+$ and $N_\eta = 4$. For each set of parameters, three different executions of the quantum circuit are performed, for different background field configurations, consequently, each point in the plot is the mean of the three individual executions and the error bars are the respective standard deviation. The solid lines represent the analytical expectations.}
    \label{fig:quark_neta_nperp16}
\end{figure*}

Even though the impact of the $N_\eta$ parameter was studied for different values of $N_\perp$, the considered values of $N_\eta$ are different for each given $N_\perp$.  The larger the $N_\perp$ and $N_\eta$m the slower and more resource consuming are the simulations. Although increasing the $N_\eta$ value leads to convergent results respecting the different $p^+$ values, the convergence for small values of $N_\eta$ makes the improvement difficult to see in \autoref{fig:quark_neta_nperp4}, \autoref{fig:quark_neta_nperp8} and \autoref{fig:quark_neta_nperp16}.
When comparing different $N_\perp$ values, one can easily infer that the larger the $N_\perp$ value, the larger the range of saturation scales that lead to results in a good agreement with the classical expectation. This result is explained by the spacing effects present in the simulations (see \autoref{fig:range_coverage_su3}).

When comparing the SU(3) results here presented with the SU(2) ones, particularly \autoref{fig:quark_neta_nperp8} with the equivalent figure in \autoref{app:SU2}, we see that the SU(3) simulations clearly yield more accurate results for higher saturation scales. 
Moreover, for the SU(2) case, the tensor and controlled methods have compatible performances, which is not verified for the SU(3) case. 
This discrepancy can be easily justified by the large problem size, the small values of the background field operator, and the consequent accumulation of numerical errors.

\subsection{Gluon}

\begin{figure}
    \centering
    \includegraphics[width=0.8\linewidth]{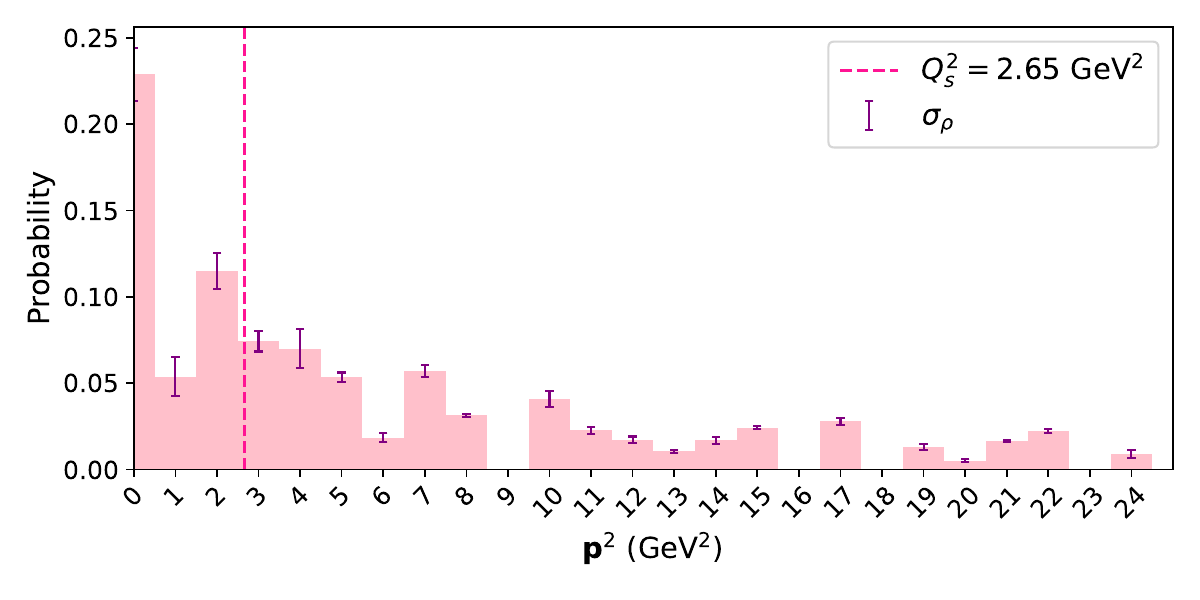}
    \caption{Squared momentum distribution for a SU(3) gluon with $N_{\perp} = 8$, $L_{\perp} = 4.8 $~GeV\textsuperscript{-1}, $m_g=0.8$~GeV, $N_{\eta} = 4$, $p^+=+\infty $~GeV, $g^2\mu = 0.5$~GeV\textsuperscript{$\frac{3}{2}$} and $N_{reps} =1 $. For this plot, three individual executions of the quantum circuit are presented, with the height of the bins being the mean of the three individual executions and the error bars being the respective standard deviation. Each execution corresponds to a different background field configuration. The vertical dashed line represents the saturation scale of that simulation.}
    \label{fig:p2gluon}
\end{figure}

The analysis of the results for the gluon case follows closely the quark case, leading to similar conclusions.
\autoref{fig:p2gluon} shows squared momentum distribution for gluon.

\begin{figure*}
    \centering
    \includegraphics[width=0.9\textwidth]{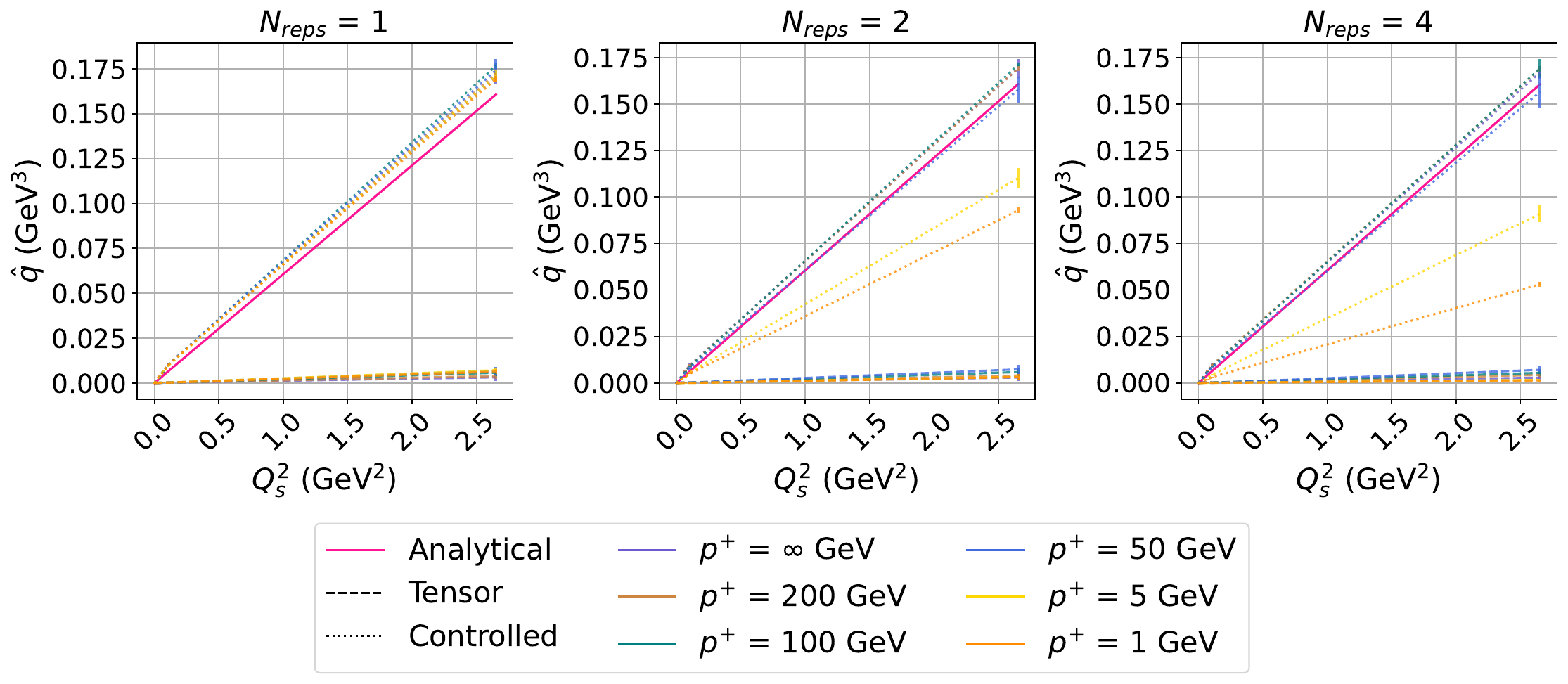}
    \caption{Jet quenching parameter $\hat{q}$ as a function of the saturation scale $Q_s^2$ for a SU(3) gluon with $N_{\perp} = 8$, $L_{\perp} = 4.8 $~GeV\textsuperscript{-1}, $m_g=0.8$, $N_{\eta} = 16$ for several values of $p^+$. For each set of parameters, three different executions of the quantum circuit are performed, for different background field configurations, consequently, each point in the plot is the mean of the three individual executions and the error bars are the respective standard deviation. The solid lines represent the analytical expectations.}
    \label{fig:gluon_reps}
\end{figure*}

The impact of $N_{reps}$ on the jet quenching parameter is shown in \autoref{fig:gluon_reps}, including comparison with the analytical expectation, \autoref{eq:qhat}. Noteworthy is that saturation scale range for which we found a good agreement with the analytical expectation is smaller than for the quark case due to more significant discrete lattice effects.

\begin{figure*}
    \centering
    \includegraphics[width=0.9\textwidth]{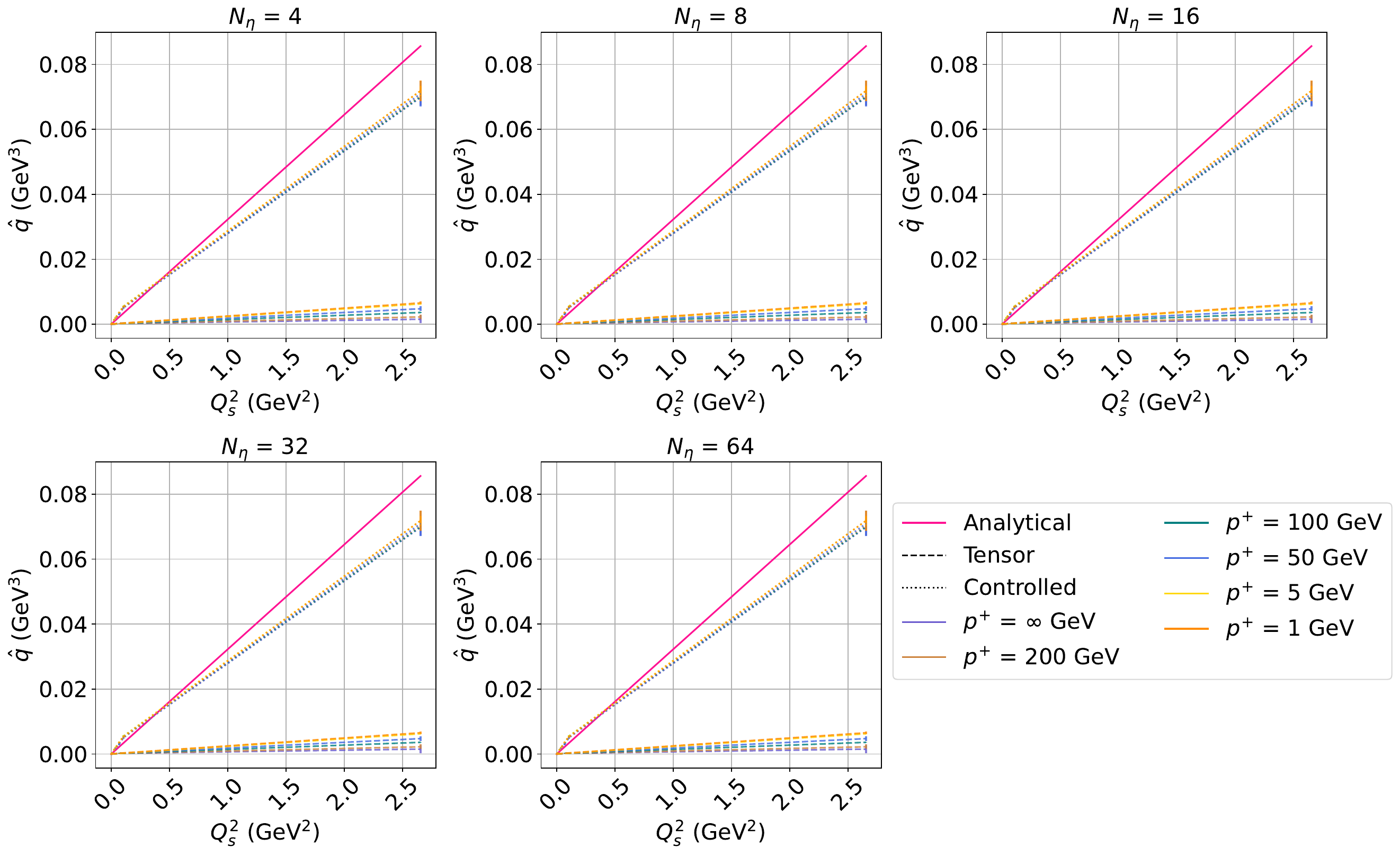}
    \caption{Jet quenching parameter $\hat{q}$ as a function of the saturation scale $Q_s^2$ for a SU(3) gluon with $N_{\perp} = 4$, $L_{\perp} = 4.8 $~GeV\textsuperscript{-1}, $N_{reps} = 1$, $m_g=0.8$~GeV for several values of $p^+$ and $N_\eta$. For each set of parameters, three different executions of the quantum circuit are performed, for different background field configurations, consequently, each point in the plot is the mean of the three individual executions and the error bars are the respective standard deviation. The solid lines represent the analytical expectations.}
    \label{fig:gluon_neta_nperp4}
\end{figure*}

\begin{figure*}
    \centering
    \includegraphics[width=0.9\textwidth]{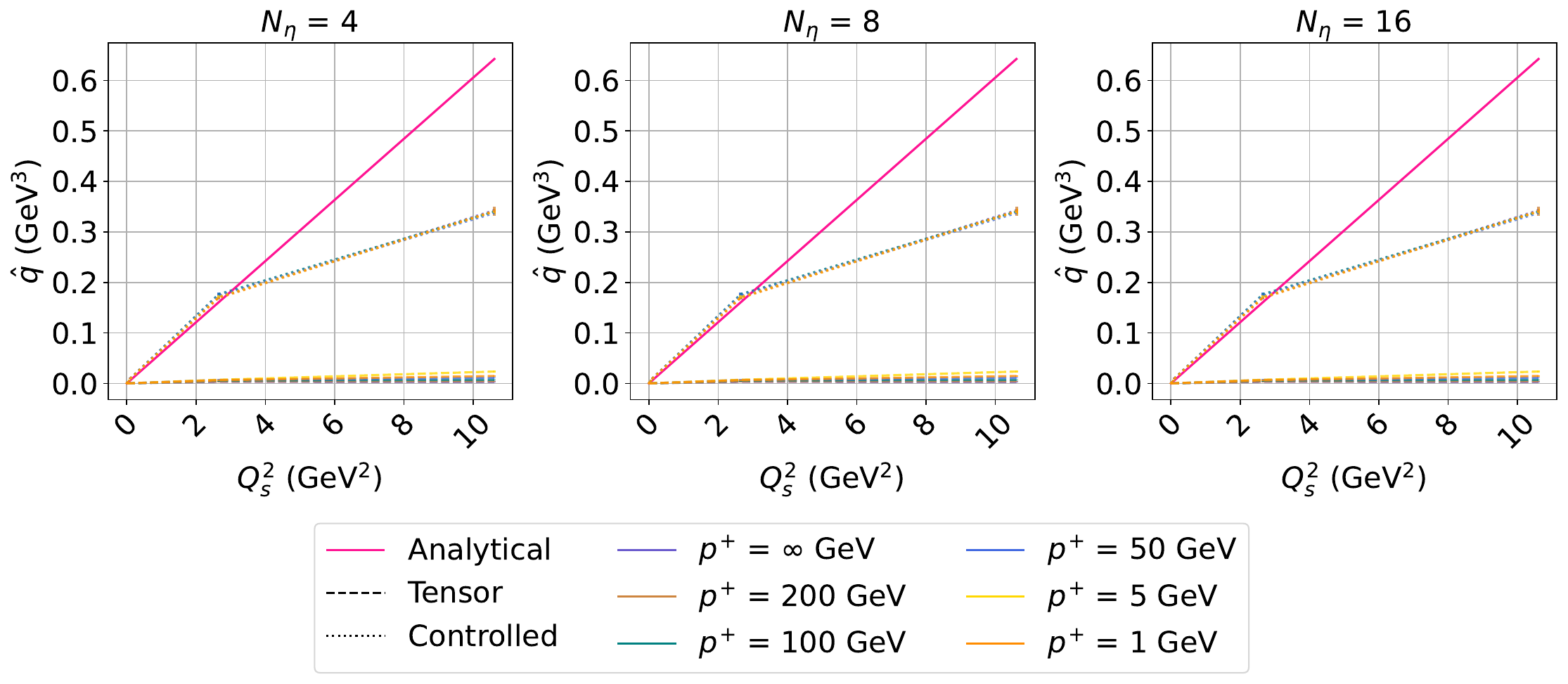}
    \caption{Jet quenching parameter $\hat{q}$ as a function of the saturation scale $Q_s^2$ for a SU(3) gluon with $N_{\perp} = 8$, $L_{\perp} = 4.8 $~GeV\textsuperscript{-1}, $N_{reps} = 1$, $m_g=0.8$~GeV for several values of $p^+$ and $N_\eta$. For each set of parameters, three different executions of the quantum circuit are performed, for different background field configurations, consequently, each point in the plot is the mean of the three individual executions and the error bars are the respective standard deviation. The solid lines represent the analytical expectations.}
    \label{fig:gluon_neta_nperp8}
\end{figure*}

\begin{figure*}
    \centering
    \includegraphics[width=0.9\textwidth]{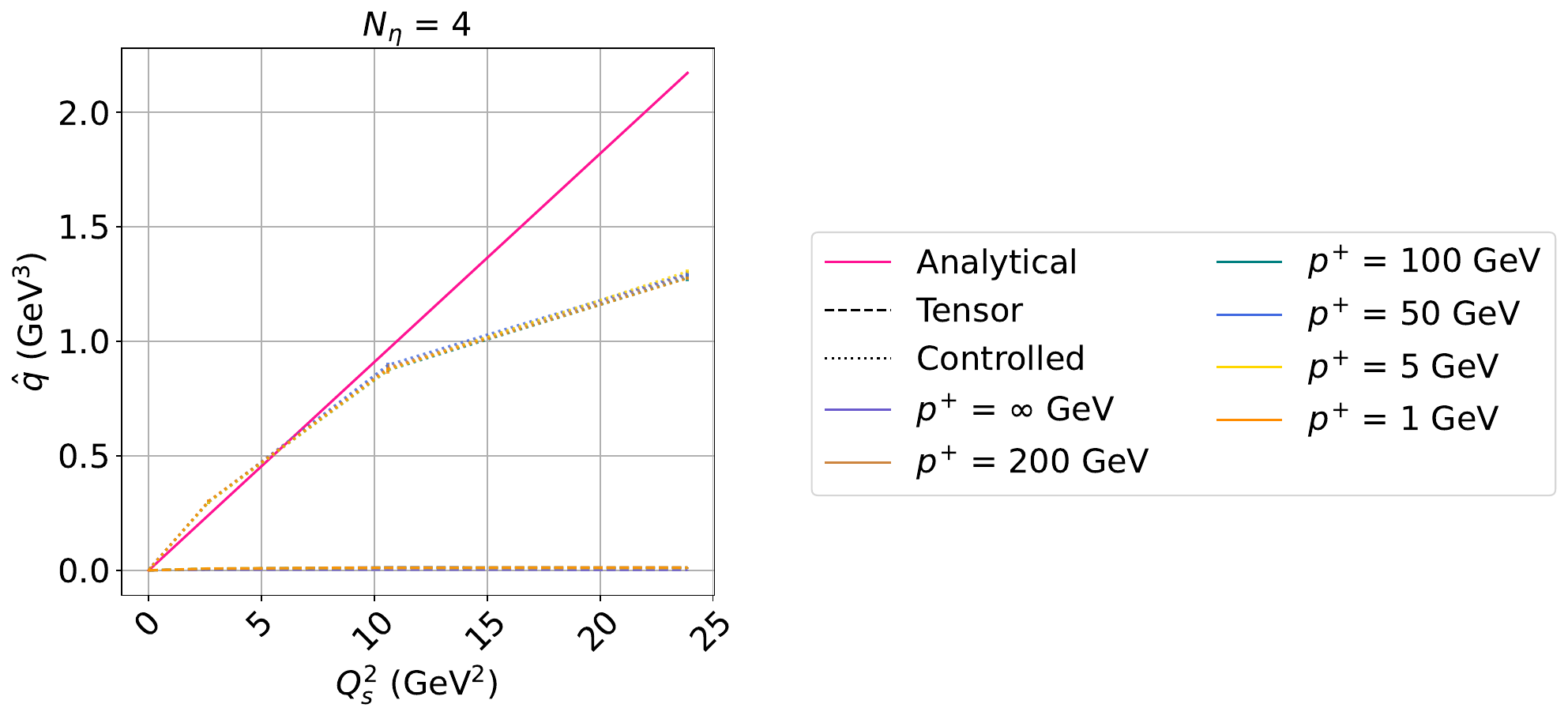}
    \caption{Jet quenching parameter $\hat{q}$ as a function of the saturation scale $Q_s^2$ for a SU(3) gluon with $N_{\perp} = 16$, $L_{\perp} = 4.8 $~GeV\textsuperscript{-1}, $N_{reps} = 1$, $m_g=0.8$~GeV for several values of $p^+$ and $N_\eta = 4$. For each set of parameters, three different executions of the quantum circuit are performed, for different background field configurations, consequently, each point in the plot is the mean of the three individual executions and the error bars are the respective standard deviation. The solid lines represent the analytical expectations.}
    \label{fig:gluon_neta_nperp16}
\end{figure*}

\autoref{fig:gluon_neta_nperp4}, \autoref{fig:gluon_neta_nperp8}, and \autoref{fig:gluon_neta_nperp16} show the impact of different choices of $N_\perp$ on the jet quenching parameter. Again, the reduced range of saturation scales for which there is agreement with the analytical baseline is explained by more significant discrete lattice effects for gluons than for quarks.

Similar to the discussion in the quark case, a comparison between the SU(3) results presented here and the SU(2) results, particularly between \autoref{fig:gluon_neta_nperp8} and the corresponding figure in \autoref{app:SU2}, shows that SU(3) and SU(2) simulations achieve comparable accuracy.
This is noteworthy given that the analytical results differ between the SU(2) and SU(3) theories. 
The discrepancy between the implementation methods is also observed in this case.

\section{Conclusions}

This work investigated the feasibility of simulating the evolution of jets in a quantum computer, more precisely, the propagation of a single SU(3) parton through a medium. The results obtained, for both the quark and gluon cases, were analised and compared against the classical expectation and discussed. 

In a quantum simulation environment, the results closely matched analytical expectations for saturation scales below approximately $30$~GeV\textsuperscript{2}, demonstrating the potential of using quantum computers to simulate and solve this class of problems. 

We found that the longitudinal evolution should be divided into as many steps as the number of divisions of the longitudinal background medium. When the longitudinal evolution is divided into more steps than the background medium ones, there is a decrease in performance for $p^+$ that do not correspond to the eikonal or sub-eikonal regimes. This convergence to smaller value than the analytical expectation for small $p^+$ is in agreement with the results in \cite{LI2023}. Furthermore, more divisions of the background field lead to better results, especially for smaller values of $p^+$. Although some preliminary tests for SU(2) partons have shown that the controlled and tensorial methods have similar performances, for the physical meaningful theory -- the SU(3) -- the performance of the tensorial method vanishes. 

For the physical meaningful SU(3) theory, we found that, for the quark case, we have better results for higher saturation scales when compared with the SU(2) analysis and that for the gluon case, the performance is compatible even though the analytical values in SU(2) and SU(3) differ. In addition, due to the large dimensions and complexity of relevant quantum simulations of both SU(2) and SU(3) theories, fault-tolerant quantum computers are needed to adequately execute them in real devices. Consequently, in the long term, there is no major advantage in running SU(2) simulations instead of the realistic SU(3).

Even though the results obtained in this work are promising, there are still many challenges to overcome. The most immediate extension of this work involves extending the Fock space to better approximate the full evolution of jets\footnote{ Although in \cite{Barata2023} the Fock space was already extended to include the gluon production, the simulation was performed for a very simple case with $N_\perp=1$. Therefore, the next major step in this direction is to extend the Fock space using lattices and saturation scales similar to those employed here.}. Another direction to follow in the future relates to the interaction with the medium. The quantum nature of the quark-gluon plasma makes the integration of the medium generation/simulation in the quantum algorithm itself a very interesting next step. Designing a procedure to simulate the entire interaction with the medium on a quantum computer would allow for a full quantum simulation of jet evolution. This full quantum regime would likely be more advantageous when fault-tolerant quantum computers become available. Furthermore, the natural representation of the SU(3) group would be a unit of information with three different states, one for each colour. From the quantum information perspective, this would involve a qutrit \cite{gokhale2020,Roy_2023} or a hybrid qubit-qutrit \cite{Bakkegaard_2019} representation. Finally, the proposed algorithm is expected to be improvable by exploring quantum error mitigation techniques \cite{Cai2023,Temme2017,van_den_Berg_2023,Wallman_2016}. 
It should be remarked, nonetheless, that even if this is a promising path, it is not foreseeable that such error mitigation techniques will drastically improve the current results in a near future, due to the high dimensionality of the circuits.

In conclusion, this work, along with other studies \cite{Barata2021,Barata2022,Barata2023}, represents one of the first steps on the long road towards simulating the full evolution of jets on a quantum computer. Even though the results demonstrate the feasibility of the quantum simulation of the jet evolution, there are still
challenges to overcome and many steps to follow to achieve a full simulation of jets in a quantum computer.

\begin{acknowledgements}
We would like to thank João Barata and Carlos Salgado for the relevant and interesting meetings and all the helpful information. We would also like to thank Ricardo Ribeiro and José Rufino for kindly providing access to some of the computing systems used in this work. We acknowledge the use of IBM Quantum services for this work. This work was produced with the computational support of INCD funded by FCT and FEDER under the project with reference 2023.10635.CPCA.A1. This work was financed by national funds through FCT ‑ Fundação para a Ciência e a Tecnologia, I.P., within the scope of the project CERN/FIS‑ PAR/0032/2021 and by European Research Council (ERC) under the European Union's Horizon 2020 research and innovation programme (Grant agreement No. 835105, YoctoLHC).
\end{acknowledgements}

\appendix

\section{colour background field}
\label{app:colour_field}

The statistics of the colour background field are described by a version of the McLerran-Venugopalan model \cite{McLerran1994,McLerran1994v2} that assumes that the colour charges of the medium are correlated white-noise statistics, just as assumed in \cite{Barata2022,Barata2023}. Hence, the colour background field is a classical stochastic field, which satisfies the reduced Yang-Mills equations, 
\begin{equation}
    \left(m_g^2-\nabla_\perp^2\right)\mathcal{A}_a^-\left(x^+,\mathbf{x}\right) = \rho_a \left(x^+,\mathbf{x}\right)\, ,
    \label{eq:ym}
\end{equation}
where $m_g$ is the gluon mass, introduced to regularise the infrared divergence, ensuring the colour neutrality of the source distribution \cite{Krasnitz2003}, and $\rho_a \left(x^+,\mathbf{x}\right)$ is the colour charge density describing the medium's energetic degrees of freedom. The colour charge density is assumed to have a Gaussian correlation function, i.e., 
\begin{equation}
    \left\langle \left\langle  \rho_a \left(x^+,\mathbf{x}\right)\rho_b \left(x^+,\mathbf{y}\right)\right\rangle \right\rangle 
    = g^2\mu^2\delta_{ab}\delta^{\left(2\right)}\left(\mathbf{x}-\mathbf{y}\right)\delta\left(x^+-y^+\right)\, ,
    \label{eq:correlation}
\end{equation}
where $\left\langle \left\langle \dots \right\rangle \right\rangle$ denotes the average over different medium configurations, $g$ is the QCD coupling constant, and $\mu$ has dimension of GeV$^\frac{3}{2}$ and can be interpreted as the medium's density of scattering centres and so determines the strength of the parton-medium interaction. In the context of high-energy scattering processes, the charge density is usually integrated over the longitudinal extension of the medium. This new quantity is known is the saturation scale $Q_s^2$ defined in \autoref{eq:saturation}.

By solving \autoref{eq:ym}, the background field can finally be expressed as
\begin{equation}
    \mathcal{A}_a^-\left(x^+,\mathbf{x}\right) = \int d^2\mathbf{y} \frac{d^2 \mathbf{k}}{\left(2\pi\right)^2} \frac{e^{-i\mathbf{k}\cdot\left(\mathbf{x}-\mathbf{y}\right)}}{m_g^2+\mathbf{k}^2}\rho_a\left(x^+,\mathbf{y}\right)\, .
    \label{eq:background_field}
\end{equation}

\section{Discrete lattice effects}
\label{subsec:lattice_effects}

The discrete nature of the lattice introduces two cutoffs, an IR cutoff $\lambda_{IR} = \frac{\pi}{L_\perp}$ and an UV cutoff $\lambda_{UV} = \frac{\pi}{\Delta_\perp}=\lambda_{IR}N_\perp$. 
To ensure the absence of \textit{spacing effects}, the lattice spacing $\Delta_\perp$ should be chosen in such a way that the momentum transfer $Q_s$ is much smaller than the UV cutoff $\lambda_{UV}$, i.e.,
\begin{equation}
    Q_s \ll \lambda_{UV} \Rightarrow Q_s \ll \frac{\pi}{\Delta_\perp} \Rightarrow \Delta_\perp \ll \frac{\pi}{Q_s}\, ,
\end{equation}
and that the physical IR regulator $m_g$ is much larger than the IR cutoff $\lambda_{IR}$, i.e.,
\begin{equation}
    m_g \gg \lambda_{IR}
    \Rightarrow m_g \gg \frac{\pi}{L_\perp} 
    \Rightarrow L_\perp \gg \frac{\pi}{m_g} \Rightarrow N_\perp \Delta_\perp \gg \frac{\pi}{m_g} \Rightarrow \Delta_\perp \gg \frac{\pi}{N_\perp m_g}\, .
\end{equation}
Jointly, these two conditions lead to
\begin{equation}
    \frac{\pi}{N_\perp m_g} \ll \Delta_\perp \ll \frac{\pi}{Q_s}\, .
    \label{eq:spacing}
\end{equation}
Due to the lattice periodicity, when there are \textit{finite size effects}, the lattice edges affect the final momentum distribution by making it asymptotically uniform. Mathematically, a uniform momentum distribution can be described as

\begin{align}
    \braket{\mathbf{p}^2} &\xrightarrow{L_\eta \gg 0} \frac{1}{\left(2N_\perp\right)^2} \sum_{k_x = -N_\perp}^{N_\perp-1} \sum_{k_y = -N_\perp}^{N_\perp-1} \left(k_x^2 + k_y^2\right) \Delta_{\perp p}^2 \nonumber\\
    &= \frac{1}{\left(2N_\perp\right)^2} \left(\frac{4}{3}N_\perp^2 \left(2N_\perp^2+1\right)\right)\Delta_{\perp p}^2  \nonumber\\
    &= \frac{2}{3}N_\perp^2\Delta_{\perp p}^2 + \frac{1}{3}\Delta_{\perp p}^2 \approx \frac{2}{3}N_\perp^2\Delta_{\perp p}^2 = \frac{2}{3}N_\perp^2\left(\frac{\pi}{L_\perp}\right)^2 \nonumber\\
    &= \frac{2}{3}N_\perp^2\left(\frac{\pi}{L_\perp}\right)^2 = \frac{2}{3}\frac{\pi^2}{\Delta_\perp^2} \equiv \braket{\textbf{p}^2}_{\text{uniform}}\, .
\end{align}
Now, defining $L_{sat} \equiv \frac{\braket{\textbf{p}^2}_{\text{uniform}}}{\hat{q}}$, to avoid the uniform momentum distribution, one should ensure a set of parameters where $L_{sat} \gg L_\eta$, i.e., the total time of the evolution should be smaller than the time needed for the lattice edges to affect the final momentum distribution. For the quark case, recalling the expressions for $Q_s^2$, \autoref{eq:saturation}, and $\hat{q}$, \autoref{eq:qhat}, this condition is rewritten as
 \begin{align}
    L_{sat} \gg L_\eta &\Rightarrow \frac{2}{3}\frac{\pi^2}{\Delta_\perp^2 \hat{q}} \gg L_\eta \nonumber \\
     &\Rightarrow \frac{2}{\sqrt{3}}\frac{\pi}{ Q_s \left(\log\left(\frac{1+\frac{\Delta_\perp^2m_g^2}{\pi^2}}{\frac{1}{N_\perp^2}+\frac{\Delta_\perp^2m_g^2}{\pi^2}}\right)-\frac{\Delta_\perp^2m_g^2}{\pi^2}\left(\frac{1}{\frac{1}{N_\perp^2}+\frac{\Delta_\perp^2m_g^2}{\pi^2}}-\frac{1}{1+\frac{\Delta_\perp^2m_g^2}{\pi^2}}\right)\right)^{\frac{1}{2}}} - \Delta_\perp \gg 0.
     \label{eq:finite_size}
 \end{align}
Consequently, to avoid the \textit{finite size effects} for a quark, one should ensure that this expression is positive.
For a gluon, the saturation scale is given by $\widetilde{Q}^2 \equiv C_A\frac{\left(g^2\mu\right)^2L_\eta}{2\pi}$ and the gluon version of \autoref{eq:finite_size} is obtained by simply replacing  $Q_s \rightarrow \widetilde{Q}$.

Spacing effects are avoided for choices of $Q_s^2$,$N_\perp$, $L_\perp$, and $m_g$, for which \autoref{eq:spacing} is satisfied.

\begin{figure*}
    \centering
    \includegraphics[width=0.9\textwidth]{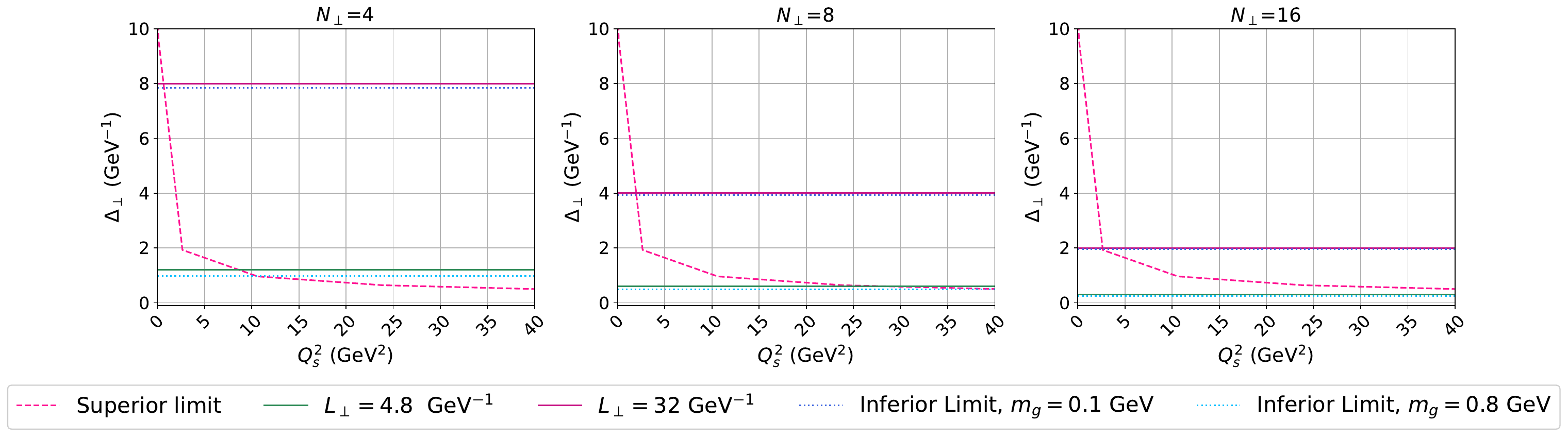}
    \caption[{Spacing effects for the SU(3) gauge group for $N_\perp \in [4,8,16]$, $L_\perp \in [4.8,32] $~GeV\textsuperscript{-1} and $m_g \in [0.1, 0.8]$~GeV for several values of $g^2\mu$.}]{Spacing effects for the SU(3) gauge group for $N_\perp \in [4,8,16]$, $L_\perp \in [4.8,32] $~GeV\textsuperscript{-1} and $m_g \in [0.1, 0.8]$~GeV for several values of $g^2\mu$. The upper limit corresponds to the condition $\Delta_\perp \ll \frac{\pi}{Q_s}$, and the lower limits corresponds to the condition $\Delta_\perp \gg \frac{\pi}{N_\perp m_g}$, one for each $m_g$ value. The filled green and pink lines correspond to the two possible values of $\Delta_\perp$ for each value of $N_\perp$.}
    \label{fig:range_coverage_su3}
\end{figure*}

\begin{figure*}
    \centering
    \includegraphics[width= 0.9\textwidth]{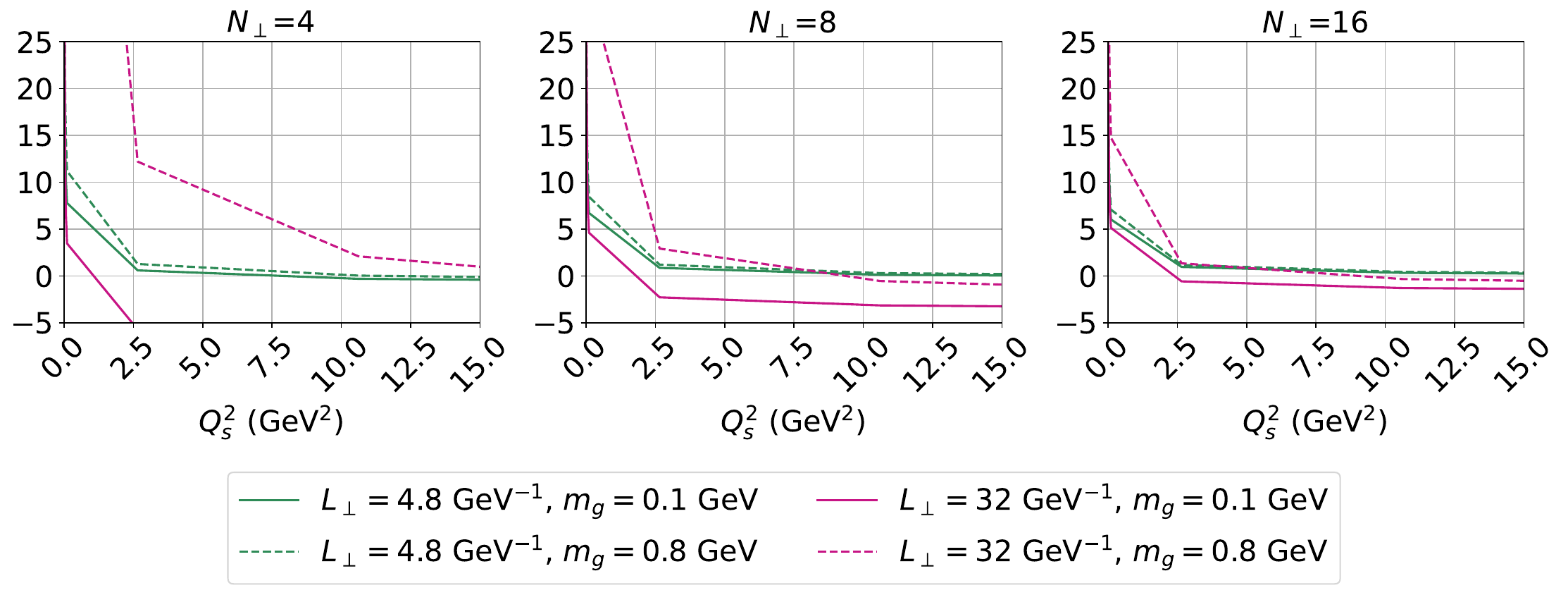}
    \caption[{Finite size effects for a SU(3) quark for $N_\perp \in [4,8,16]$, $L_\perp \in [4.8,32] $~GeV\textsuperscript{-1} and $m_g \in [0.1, 0.8] $~GeV for several values of $g^2\mu$.}]{Finite size effects for a SU(3) quark for $N_\perp \in [4,8,16]$, $L_\perp \in [4.8,32] $~GeV\textsuperscript{-1} and $m_g \in [0.1, 0.8] $~GeV for several values of $g^2\mu$. Each curve corresponds to the condition in \autoref{eq:finite_size} for the two possible values of $\Delta_\perp$ for each value of $N_\perp$, which are related to the $L_\perp$ values.}
    \label{fig:q2broadning_coverage_su3}
\end{figure*}

\begin{figure*}
    \centering
    \includegraphics[width=0.9\textwidth]{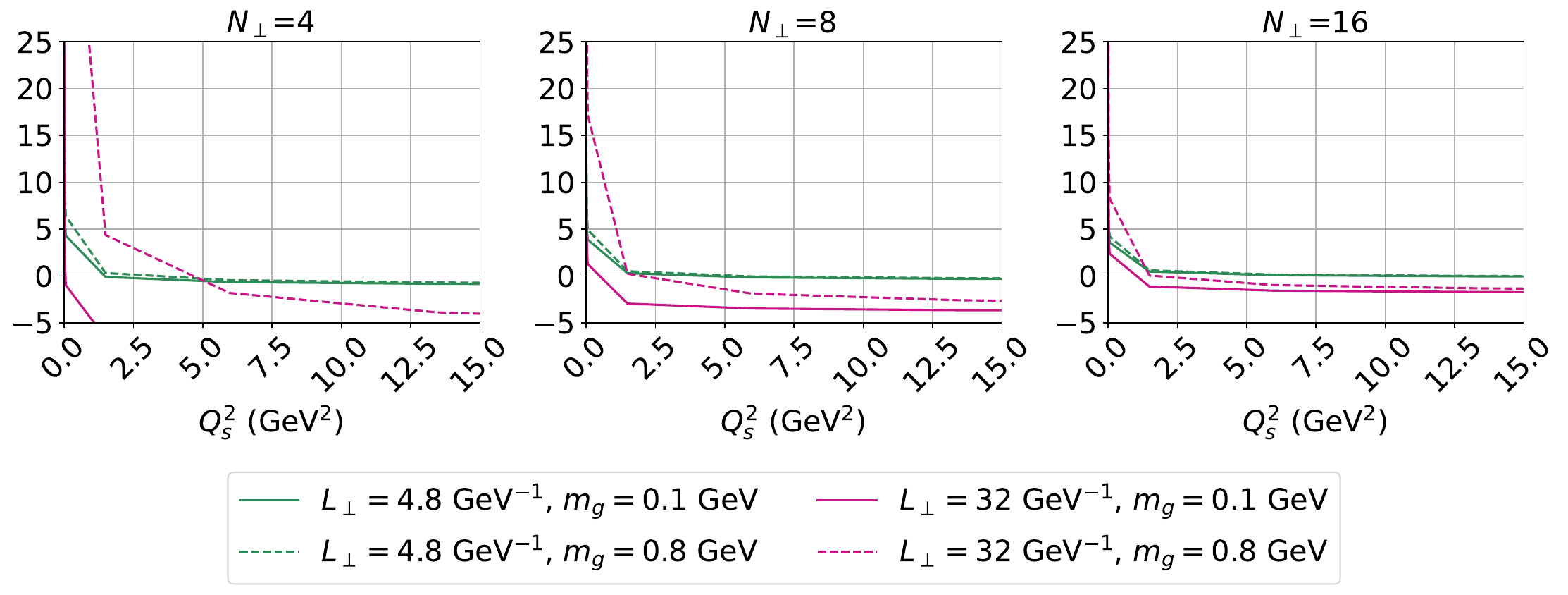}
    \caption[{Finite size effects for a SU(3) gluon for $N_\perp \in [4,8,16]$, $L_\perp \in [4.8,32] $~GeV\textsuperscript{-1} and $m_g \in [0.1, 0.8] $~GeV for several values of $g^2\mu$.}]{Finite size effects for a SU(3) gluon for $N_\perp \in [4,8,16]$, $L_\perp \in [4.8,32] $~GeV\textsuperscript{-1} and $m_g \in [0.1, 0.8] $~GeV for several values of $g^2\mu$. 
    }
    \label{fig:gluon_q2broadning_coverage_su3}
\end{figure*}

\autoref{fig:range_coverage_su3} shows the conditions regarding spacing effects, which are the same for quarks and gluons, for different choices of $Q_s^2$, $N_\perp$, $L_\perp$, and $m_g$. The conditions for finite size effects are shown, separately for quarks and gluons, in \autoref{fig:q2broadning_coverage_su3} and \autoref{fig:gluon_q2broadning_coverage_su3}.

The most salient feature in \autoref{fig:range_coverage_su3} is that for $m_g = 0.1 $~GeV, using $4.8 $~GeV\textsuperscript{-1} for $L_\perp$, leads to spacing effects for all values of $g^2\mu$ and $N_\perp$. This is the reason why when $L_\perp = 4.8 $~GeV\textsuperscript{-1}, $m_g = 0.8 $~GeV is chosen, and when $L_\perp = 32 $~GeV\textsuperscript{-1}, $m_g = 0.1 $~GeV is chosen. Concerning the upper limit, choosing $L_\perp = 4.8 $~GeV\textsuperscript{-1} and $m_g = 0.8 $~GeV always allows a larger range of saturation scales that are not affected by spacing effects than choosing $L_\perp = 32 $~GeV\textsuperscript{-1}, $m_g = 0.1 $. Consequently, the former set of parameters will be used.

For the quark case, the larger the values of $N_\perp$, the smaller the values of $Q_s^2$ that lead to the finite effects, and the larger the values of $N_\perp$ the larger the values of $Q_s^2$ that lead to the absence of the finite size effects. For $L_\perp = 32 $~GeV\textsuperscript{-1}, the values become ``more negative'' than for $L_\perp = 4.8 $~GeV\textsuperscript{-1}, which converge to a small value around $0$ as $Q_s^2$ increases. Furthermore, for $m_g = 0.1 $~GeV the lines tend to become negative for smaller values of $Q_s^2$ than for $m_g = 0.8 $~GeV, which is more evident for smaller values of $N_\perp$.

When looking at the results in \autoref{fig:gluon_q2broadning_coverage_su3}, one can see that the morphology of the curves is quite similar to the ones for the quark case, in \autoref{fig:q2broadning_coverage_su3}, with the main and most significant different being the fact that the curves become negative for smaller values of $Q_s^2$ than for the quark case. Consequently, one can expect that the results found for the gluon case are more affected by the finite size effects than the ones for the quark case, especially for the higher values of $Q_s^2$.

\section{Analytical jet quenching parameter}
\label{app:analytical_jet_quenching}
In a similar way to Refs. \cite{LI2023,Barata2022}, the first step to compute the analytical jet quenching parameter is to rewrite \autoref{eq:jet_quenching_par} as
\begin{equation}
    \hat{q} = \frac{\langle\langle \bra{\psi_0} U^\dagger\left(L_\eta;0\right) \hat{\mathbf{p}}^2 U\left(L_\eta;0\right)\ket{\psi_0}\rangle\rangle}{L_\eta},
\end{equation}
where $\ket{\psi_0}$ is the initial state of the system, and $U\left(L_\eta;0\right)$ is the usual time evolution operator from the light-front time $0$ to the light-front time $L_\eta$, i.e., for the whole medium. Through a Fourier transform, the $\hat{q}$ in coordinate space is given by
\begin{equation}
    \hat{q} = \frac{1}{L_\eta}\int d\mathbf{x} d\mathbf{y} d\mathbf{k} e^{-i\mathbf{k}\left(\mathbf{x}-\mathbf{y}\right)}\mathbf{k}^2\langle\langle \mathcal{W}^\dagger\left(\mathbf{y}\right)\mathcal{W}\left(\mathbf{x}\right)\rangle\rangle 
    = \frac{1}{L_\eta}\int d\mathbf{x} \langle\langle \nabla_\mathbf{y}\mathcal{W}^\dagger\left(\mathbf{y}\right)\cdot \nabla_\mathbf{x}\mathcal{W}\left(\mathbf{x}\right)\rangle\rangle\, ,
\end{equation}
where $\nabla_\mathbf{k}$ is the directional derivative along $\mathbf{k}$, and $\mathcal{W}$ is a Wilson line along $x^+$ in the light-cone framework which can be explicitly written as
\begin{equation}
    \mathcal{W}\left(\mathbf{x}\right) = e^{-ig\int_0^{L_\eta} dx^+ \mathcal{A}^-\left(x^+,\mathbf{x}\right)}\, ,
    \label{eq:wilson_line}
\end{equation}
and corresponds to the evolution operator (\autoref{eq:time_evolution}) in the eikonal limit, i.e., $p^+ \rightarrow \infty$ \footnote{In \cite{Blaizot2013} is shown that assuming finite $p^+$ values, and so, including the kinetic term at the leading eikonal order, does not affect the $\hat{q}$.}. For the quark case, the above Wilson line should be used in the fundamental form, i.e.,  $\mathcal{A}^-\left(x^+,\mathbf{x}\right)$ should be replaced by $\mathcal{A}_a^-\left(x^+,\mathbf{x}\right)\mathbf{t}^a$, and for the gluon case, the adjoint form, i.e., replaced by $\mathcal{A}_a^-\left(x^+,\mathbf{x}\right){\mathbf{T}'^a}$. 

By choosing coincident coordinates $\mathbf{y} = \mathbf{x}$, and substituting the Wilson line (\autoref{eq:wilson_line}) in the expression for $\hat{q}$, one obtains

\begin{align}
    \hat{q} &= \frac{1}{L_\eta} \langle\langle \nabla_\mathbf{y}\mathcal{W}^\dagger(\mathbf{y})\cdot \nabla_\mathbf{x}\mathcal{W}(\mathbf{x})\rangle\rangle _{\mathbf{y=x}} \nonumber \\
     &= \frac{1}{L_\eta} \langle\langle \nabla_\mathbf{y}e^{ig\int_0^{L_\eta} dy^+ \mathcal{A}^{-\dagger}(y^+,\mathbf{y})}\cdot \nabla_\mathbf{x}e^{-ig\int_0^{L_\eta} dx^+ \mathcal{A}^-(x^+,\mathbf{x})}\rangle\rangle _{\mathbf{y=x}} \nonumber \\
    &= \frac{g^2}{L_\eta} \int_0^{L_\eta} dx^+ \int_0^{L_\eta} dy^+ 
    \cdot\langle\langle {\nabla_\mathbf{y}\mathcal{A}_a^{-\dagger}(y^+,\mathbf{y})t^a}\cdot \nabla_\mathbf{x} \mathcal{A}_b^-(x^+,\mathbf{x})t^b\rangle\rangle _{\mathbf{y=x}}.
\end{align}

In \autoref{eq:wilson_line}, the Wilson line is explicitly written in the fundamental form for the quark case, and, if one wants to analyze the gluon case, one should replace the fundamental matrices with the adjoint ones. Now, by using the background field definition (\autoref{eq:background_field}) and by omitting the colour indices for simplicity, the $\hat{q}$ for a quark can be rewritten as

\begin{multline}
    \hat{q} = \frac{g^2}{L_\eta}\int_0^{L_\eta} dx^+ \int_0^{L_\eta} dy^+
    \\
    \langle\langle { \nabla_\mathbf{y}\left(\int d\mathbf{z}\frac{d^2\mathbf{k}}{\left(2\pi\right)^2}\frac{e^{i\mathbf{k}\left(\mathbf{y}-\mathbf{z}\right)}}{m_g^2+\mathbf{k^2}}\rho\left(y^+,\mathbf{z}\right)t\right)}\cdot \nabla_\mathbf{x}\left(\int d\mathbf{w} \frac{d^2\mathbf{l}}{\left(2\pi\right)^2}\frac{e^{-i\mathbf{l}\left(\mathbf{x}-\mathbf{w}\right)}}{m_g^2+\mathbf{l^2}}\rho\left(x^+,\mathbf{w}\right)t\right)\rangle\rangle _{\mathbf{y=x}} \\
    = \frac{g^2}{L_\eta}\int_0^{L_\eta} dx^+ \int_0^{L_\eta} dy^+\int d\mathbf{z}\frac{d^2\mathbf{k}}{\left(2\pi\right)^2} 
    \\\cdot \mathbf{k} \frac{e^{i\mathbf{k}\left(\mathbf{x}-\mathbf{z}\right)}}{m_g^2+\mathbf{k^2}}
    \cdot \int d\mathbf{w} \frac{d^2\mathbf{l}}{\left(2\pi\right)^2} \mathbf{l}\frac{e^{-i\mathbf{l}\left(\mathbf{x}-\mathbf{w}\right)}}{m_g^2+\mathbf{l^2}}\langle\langle\rho\left(x^+,\mathbf{z}\right)\rho\left(y^+,\mathbf{w}\right)\rangle\rangle t\cdot t. 
\end{multline}

Now, remembering the colour charge density correlator (\autoref{eq:correlation}), one can write
\begin{equation}
\begin{aligned}
    \langle\langle\rho\left(x^+,\mathbf{z}\right)\rho\left(y^+,\mathbf{w}\right)\rangle\rangle t\cdot t &\equiv \sum_a \sum_b \langle\langle\rho_a\left(x^+,\mathbf{z}\right)\rho_b\left(y^+,\mathbf{w}\right)\rangle\rangle t^a\cdot t^b  \\
    &= \sum_a \sum_b g^2\mu^2\delta_{ab}\delta\left(x^+-y^+\right)\delta\left(\mathbf{z}-\mathbf{w}\right)C_F \hat{I}.
\end{aligned}
\label{eq:quark_corr}
\end{equation}

For the gluon case, the $\mathbf{t}$ matrices should be replaced by the ${\mathbf{T}'}$ matrices, i.e.,
\begin{equation}
\begin{aligned}
    \langle\langle\rho\left(x^+,\mathbf{z}\right)\rho\left(y^+,\mathbf{w}\right)\rangle\rangle {\mathbf{T}'}\cdot {\mathbf{T}'} &\equiv \sum_a \sum_b g^2\mu^2\delta_{ab}\delta\left(x^+-y^+\right)\delta\left(\mathbf{z}-\mathbf{w}\right)C_A \hat{I}.
\end{aligned}
\label{eq:gluon_corr}
\end{equation}

Finally, including the result in \autoref{eq:quark_corr} in the expression for $\hat{q}$, one obtains
\begin{equation}
\begin{aligned}
    \hat{q} &= \frac{g^2}{L_\eta} g^2\mu^2 C_F\int_0^{L_\eta} dx^+ \int d\mathbf{z}\frac{d^2\mathbf{k}}{\left(2\pi\right)^2} \mathbf{k} \frac{e^{i\mathbf{k}\left(\mathbf{x}-\mathbf{z}\right)}}{m_g^2+\mathbf{k^2}} \int \frac{d^2\mathbf{l}}{\left(2\pi\right)^2} \mathbf{l}\frac{e^{-i\mathbf{l}\left(\mathbf{x}-\mathbf{z}\right)}}{m_g^2+\mathbf{l^2}} \\
    & =  g^4\mu^2 C_F\int \frac{d^2\mathbf{p}_\perp}{\left(2\pi\right)^2} \frac{\mathbf{p}_\perp^2}{\left(m_g^2+\mathbf{p}_\perp^2\right)^2}
\end{aligned}
\end{equation}

The discrete nature of the transverse lattice introduces two momentum cutoffs, which are the Infra-Red (IR) and the Ultra-violet (UV) cutoffs, and which are defined as $\lambda_{IR} = \frac{\pi}{L_\perp}$ and $\lambda_{UV} = \frac{\pi}{\Delta_\perp}=\lambda_{IR}N_\perp$, respectively. Thus, the above integral can be rewritten in polar coordinates, with $p$ the module of $\mathbf{p}_\perp$, as
\begin{equation}
\begin{aligned}
    \hat{q} &=  g^4\mu^2 C_F\int_{\lambda_{IR}}^{{\lambda_{UV}}} dp\int_{0}^{2\pi} \frac{d\theta}{\left(2\pi\right)^2}p \frac{p^2}{\left(m_g^2+p^2\right)^2} \\
    &=  g^4\mu^2 C_F\int_{\lambda_{IR}}^{{\lambda_{UV}}} \frac{dp}{2\pi} \frac{p^3}{\left(m_g^2+p^2\right)^2} \\
    &=  \frac{g^4\mu^2 C_F}{4\pi}\left(\log\left(\frac{1+\frac{\Delta_\perp^2m_g^2}{\pi^2}}{\frac{1}{N_\perp^2}+\frac{\Delta_\perp^2m_g^2}{\pi^2}}\right)-\frac{\Delta_\perp^2m_g^2}{\pi^2}\left(\frac{1}{\frac{1}{N_\perp^2}+\frac{\Delta_\perp^2m_g^2}{\pi^2}}-\frac{1}{1+\frac{\Delta_\perp^2m_g^2}{\pi^2}}\right)\right).
\end{aligned}
\label{eq:qhat_analytical} 
\end{equation}

From \autoref{eq:quark_corr} and \autoref{eq:gluon_corr}, one can trivially see that when the propagating parton is a gluon instead of a quark, the factor $C_F$ should be replaced by $C_A$.

\section{SU(2) parton propagation}
\label{app:SU2}

\begin{figure*}
    \centering
    \includegraphics[width=0.9\textwidth]{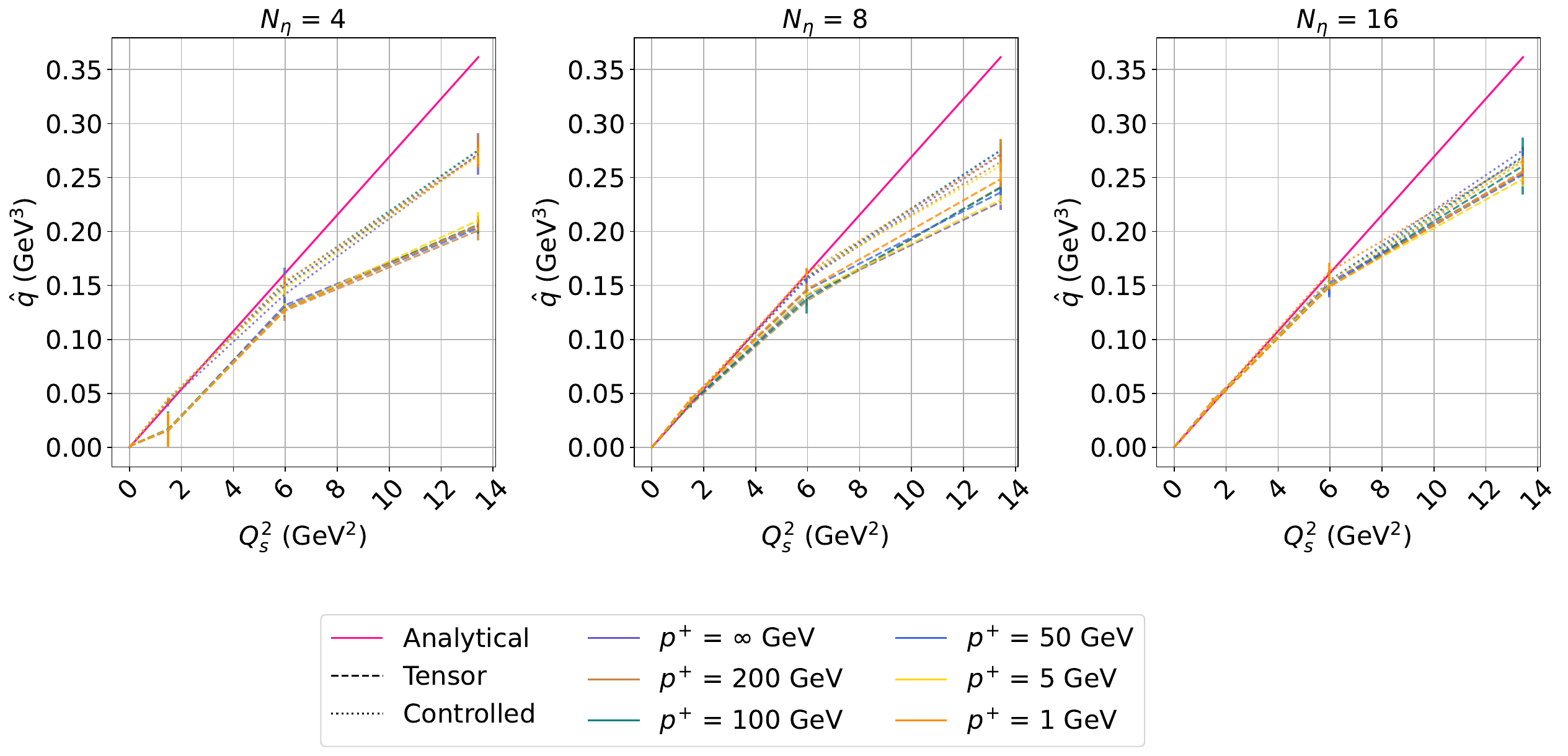}
    \caption{Jet quenching parameter $\hat{q}$ as a function of the saturation scale $Q_s^2$ for a SU(2) quark with $N_{\perp} = 8$, $L_{\perp} = 4.8 $~GeV\textsuperscript{-1}, $N_{reps} = 1$, $m_g=0.8$~GeV for several values of $p^+$ and $N_\eta$. For each set of parameters, three different executions of the quantum circuit are performed, for different background field configurations, consequently, each point in the plot is the mean of the three individual executions and the error bars are the respective standard deviation. The solid lines represent the analytical expectations.}
    \label{fig:su2quark_neta_nperp8}
\end{figure*}

\begin{figure*}
    \centering
    \includegraphics[width=0.9\textwidth]{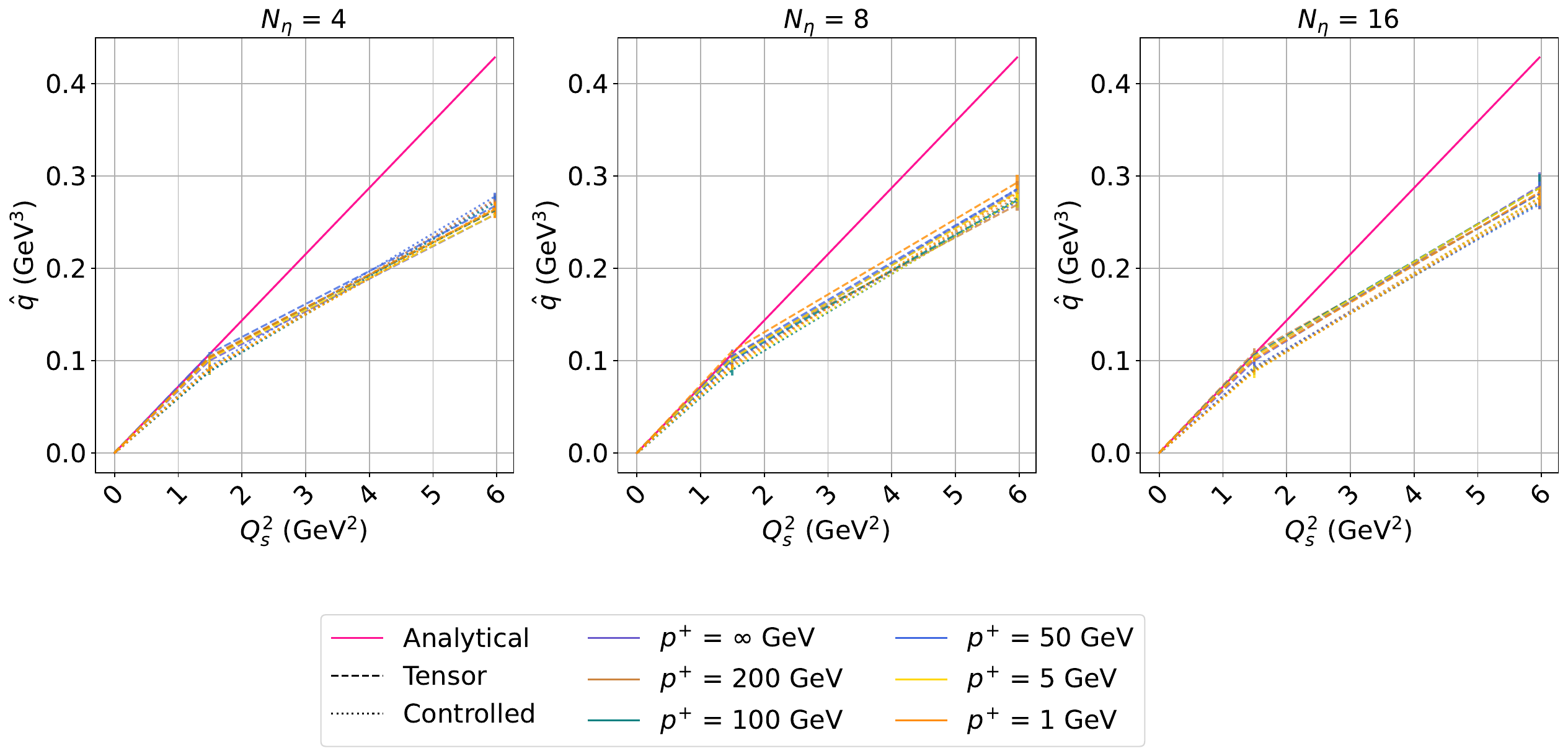}
    \caption{Jet quenching parameter $\hat{q}$ as a function of the saturation scale $Q_s^2$ for a SU(2) gluon with $N_{\perp} = 8$, $L_{\perp} = 4.8 $~GeV\textsuperscript{-1}, $N_{reps} = 1$, $m_g=0.8$~GeV for several values of $p^+$ and $N_\eta$. For each set of parameters, three different executions of the quantum circuit are performed, for different background field configurations, consequently, each point in the plot is the mean of the three individual executions and the error bars are the respective standard deviation. The solid lines represent the analytical expectations.}
    \label{fig:su2gluon_neta_nperp8}
\end{figure*}

Similarly to \autoref{fig:quark_neta_nperp8} and \autoref{fig:gluon_neta_nperp8}, in \autoref{fig:su2quark_neta_nperp8} and \autoref{fig:su2gluon_neta_nperp8} we show the impact of the $N_\eta$ parameter for a SU(2) quark and gluon, respectively, instead of the physical meaningful SU(3) partons.

\bibliographystyle{spphys}       
\bibliography{dissertation.bib}   

\end{document}